\documentclass[pra,aps,groupedaddress,twocolumn,floatfix,nofootinbib]{revtex4-1}

\usepackage{amsmath,amsthm,amsfonts,amssymb,graphics,graphicx,epsfig,color,times,natbib,subfigure,array,colortbl}
\usepackage{bbm} 
\usepackage{bm}

\newcommand{\ket}[1]{\mbox{$ | #1 \rangle $}}
\newcommand{\bra}[1]{\mbox{$ \langle #1 | $}}

\newcommand{\be}{\begin{equation}}
\newcommand{\ee}{\end{equation}}
\newcommand{\ba}{\begin{eqnarray}}
\newcommand{\ea}{\end{eqnarray}}

\newcommand{\demi}{\frac{1}{2}}

\newcommand{\one}{\leavevmode\hbox{\small1\normalsize\kern-.33em1}}

\newcommand{\moy}[1]{\langle #1 \rangle}

\def\sign{\mathrm{sign}}

\definecolor{nblue}{rgb}{0.2,0.2,0.7}
\definecolor{ngreen}{rgb}{0.2,0.6,0.2}
\definecolor{nred}{rgb}{0.8,0.2,0.2}
\definecolor{nblack}{rgb}{0,0,0}

\begin{document}

\title{Bilocal versus non-bilocal correlations in entanglement swapping experiments}

\author{Cyril Branciard$^1$, Denis Rosset$^2$,  Nicolas Gisin$^2$ and Stefano Pironio$^3$}
\affiliation{$^1$School of Mathematics and Physics, The University of Queensland, St Lucia, QLD 4072, Australia \\
$^2$Group of Applied Physics, University of Geneva, 20 rue de l'Ecole-de-M\'edecine, CH-1211 Geneva 4, Switzerland \\
$^3$Laboratoire d'Information Quantique, Universit\'e Libre de Bruxelles, 1050 Bruxelles, Belgium}

\date{\today}

\begin{abstract}

Entanglement swapping is a process by which two initially independent quantum systems can become entangled and generate nonlocal correlations. To characterize such correlations, we compare them to those predicted by {\it bilocal} models, where systems that are initially independent are described by uncorrelated states. We extend in this paper the analysis of bilocal correlations initiated in [Phys. Rev. Lett. 104, 170401 (2010)]. In particular, we derive new Bell-type inequalities based on the bilocality assumption in different scenarios, we study their possible quantum violations, and analyze their resistance to experimental imperfections. The bilocality assumption, being stronger than Bell's standard local causality assumption, lowers the requirements for the demonstration of quantumness in entanglement swapping experiments.

\end{abstract}

\maketitle

\section{Introduction}

The study of correlations between the outcomes of measurements performed on several quantum systems has led to remarkable progresses, both on fundamental aspects of quantum theory and on potential applications in quantum information technologies. Particularly intriguing are the cases where the various quantum systems are all at a distance from each other. From a fundamental point of view this situation led to the discovery of quantum nonlocality, that is, of the existence of correlations that cannot be described by a locally causal model~\cite{bell}. From an applied point of view, these studies led, quite recently, to the understanding of the power of nonlocal correlations for quantum information processing, in particular for reducing communication complexity~\cite{cleve_buhrman_cc}, for Quantum Key Distribution (QKD)~\cite{mayers-yao,DI_PRL}, private randomness generation~\cite{Pironio_randomness_Nature,colbeck_randomness}, or device independent entanglement witnesses~\cite{Bancal_DIEW}. Interestingly, in such examples nonlocal correlations can be exploited directly, in a device independent manner, independently of the Hilbert space machinery of the quantum theory. This applied side led, in turn, to a better understanding of some fundamental aspects of quantum theory and quantum information, such as for instance hidden assumptions in the abstract security analyses of QKD~\cite{acin_gisin_masanes}.

The usual starting point in such works on nonlocality are limitations---such as, e.g., Bell inequalities~\cite{bell64}---on the possible correlations between the measurement results on distant systems, following from the principle of local causality. Formally, the different systems measured in the experiment are considered to be all in an initial joint ``hidden"\footnote{We keep the terminology ``hidden'' to describe the states $\lambda$ for historical reasons~\cite{bell}. Note that these states $\lambda$ need not actually be hidden, i.e., inaccessible to the observer.} state $\lambda$, where $\lambda$ is arbitrary and could even describe the state of the entire universe prior to the measurement choices. The measurement outcome of any particular system can depend arbitrarily on the global state $\lambda$ and on the type of measurement performed on that system, but not on the measurements performed on distant systems. This last condition is Bell's local causality assumption~\cite{bell} (or Bell's ``locality assumption'', simply), which implies, e.g. in the case of three parties, that the measurement outcome probabilities can be written as
\ba
P(a,\!b,\!c|x,\!y,\!z) = \! \int \!\! \mathrm{d} \lambda \, \rho(\lambda) \, P(a|x,\!\lambda)P(b|y,\!\lambda)P(c|z,\!\lambda) , \quad \ \label{eq_locality}
\ea
where $x,y,z$ denote the measurement settings (``inputs'') chosen by the three parties, $a,b,c$ denote the corresponding measurement outcomes (``outputs''), where $\rho(\lambda)$ is a probability distribution over the set of all possible joint hidden states $\lambda$, and where
it has implicitely been assumed that the measurement choices $x,y,z$ are independent of $\lambda$.

\begin{figure}
\begin{center}
\epsfxsize=8cm
\epsfbox{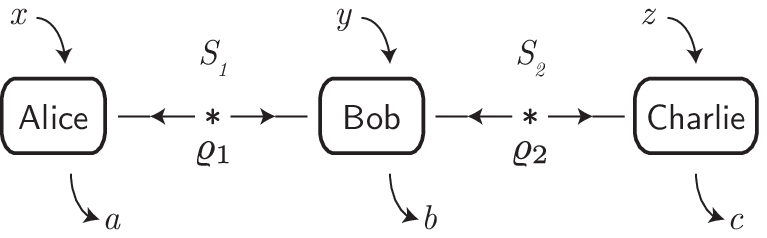}
\vspace{-.3cm}
\caption{Typical entanglement swapping scenario where three parties, Alice, Bob and Charlie, share two sources $S_1$ and $S_2$ that each emit independent pairs of particles in some quantum states $\varrho_1$ and $\varrho_2$. Bob performs a joint measurement $y$ on the two particles he receives from each source and obtains an output $b$. Depending on Bob's outcome, Alice and Charlie's systems end up in one out of different possible entangled states. Alice and Charlie apply some measurements $x$ and $z$ on their particle and obtain outputs $a$ and $c$. Such an experiment is characterized by a joint probability distribution $P(a,\!b,\!c|x,\!y,\!z)$.}
\label{fig_scenarioq}
\vspace{.3cm}
\epsfxsize=8cm
\epsfbox{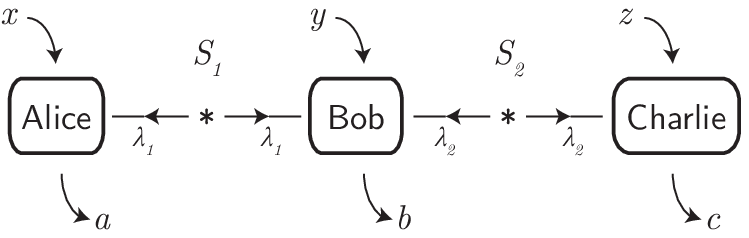}
\vspace{-.3cm}
\caption{The natural counterpart of the entanglement swapping scenario of Fig.~\ref{fig_scenarioq} in terms of a locally causal model with two independent sources of hidden states: the systems produced by the source $S_1$ are characterized by hidden states $\lambda_1$, while those from the source $S_2$ are characterized by hidden states $\lambda_2$. The two sources are assumed to be independent, hence the joint distribution $\rho(\lambda_1,\lambda_2)$ of hidden states has the product form $\rho(\lambda_1,\lambda_2)=\rho_1(\lambda_1)\rho_2(\lambda_2)$, as in eq.~(\ref{eq_bilocality}).}
\label{fig_scenariol}
\end{center}
\end{figure}

Nowadays, fast progress towards advanced demonstrations of quantum communication networks, involving quantum repeaters~\cite{quantum_repeaters} based on entanglement swappings~\cite{zukowski_event-ready-detectors_1993} and quantum memories~\cite{quantum_memories}, are underway in many labs around the world. In these future quantum networks, several independent sources of entangled qubit pairs will distribute entanglement to partners who will then connect their neighbours by performing joint measurements on two (or more) qubits, each entangled with one neighbouring qubit, as illustrated for the simple case of three partners in Fig.~\ref{fig_scenarioq}.
Such experiments have an interesting feature that has so far received little attention in previous works on nonlocality: the multipartite correlations between the measurement results at each site do not originate from a single multipartite entangled state, but from a series of bipartite entangled states that are initially independent and uncorrelated from each other; i.e., there is not a unique inital joint state (the analogue of $\lambda$ in a locally causal model) that is responsible for the observed correlations, but these are instead created from smaller systems through joint measurements. 

To understand and characterize the nonlocal properties exhibited in such experiments, it is natural to compare them to models where independent systems are characterized by different, uncorrelated hidden states $\lambda$. In the case, e.g., of the experiment of Figure~\ref{fig_scenarioq}, one would thus replace Bell's locality condition~(\ref{eq_locality}) by
\ba
P(a,\!b,\!c|x,\!y,\!z) &\!=\!& \int\!\!\!\!\!\int \! \mathrm{d} \lambda_1 \mathrm{d} \lambda_2 \, \rho_1(\lambda_1) \, \rho_2(\lambda_2) \nonumber \\
&& \quad \ P(a|x,\!\lambda_1)P(b|y,\!\lambda_1,\!\lambda_2)P(c|z,\!\lambda_2) , \quad \ \ \label{eq_bilocality}
\ea
where $\lambda_1$ characterizes the joint state of the systems produced by the source $S_1$ and $\lambda_2$ for the source $S_2$; see Figure~\ref{fig_scenariol}.
Of course, one can never exclude on pure logical grounds that systems that appear independent to us, such as pairs of particles produced by different sources, are not in fact correlated in some hidden way. But, quoting Bell, ``this way of arranging quantum mechanical correlations would be even more mind-boggling than one in which causal chains go faster than light. Apparently separate parts of the world would be deeply and conspiratorially entangled"~\cite{bell}.

Motivated by the earlier works~\cite{gisin_gisin_02,GHZZ}, the study of correlations between the results of measurements performed in quantum networks was initiated in a recent letter~\cite{biloc1_PRL} from the point of view just introduced. This leads to interesting new scenarios: because of the assumption that independent sources are characterized by different and independent $\lambda$'s, this lowers the requirements on experiments for the demonstration of quantumness of a network and, to start with the simplest case, the demonstration of quantumness of an entanglement swapping process. Such studies may lead to new applications, in the spirit of device independent quantum information processing~\cite{mayers-yao,DI_PRL,Pironio_randomness_Nature,colbeck_randomness,Bancal_DIEW}. 

The general approach considered here should also contribute to the characterization of the nonlocal properties associated to joint measurements, a question that has received little attention in traditional works on nonlocality (see however Refs.~\cite{vertesi_navascues,rabelo}).
Indeed, the joint measurements needed to connect neighbouring qubits in quantum networks are necessarily entangling, i.e., entanglement between remote parties appears through joint measurements and not from a joint state of distant systems---formally, the eigenvectors of the operators that describe joint measurement are entangled. Recall that a joint measurement, the so-called Bell state analyzer, is also at the core of the celebrated quantum teleportation protocol~\cite{Q_telep}. Entanglement, the characteristic property of quantum mechanics in Schr\"odinger's words~\cite{Schroedinger_1935}, thus plays a dual role, once allowing joint states of several systems and once allowing joint measurements. 

Before discussing in more details the results presented in this paper, let us finally stress how natural the assumption of independent $\lambda$'s is. Actually, an equivalent assumption is already implicit in all standard tests of Bell inequalities. Indeed, in such tests one needs to assume that the measurement settings are random and independent of the entanglement source~\cite{bell}; this is achieved, e.g., by having a local Quantum Random Number Generator (QRNG) determining the random settings~\cite{weihs}. But this makes sense only if one assumes that the sources in the QRNG are independent of the entanglement source and that they are characterized by independent and uncorrelated $\lambda$'s. Consequently, our assumption is actually not new, but merely formalizes a usually tacit assumption and extends its scope to more advanced topologies of quantum networks.

\subsection*{Structure of the paper}

In this article, we develop and formalize the approach introduced above for the simplest case: that of three partners on a line, with two independent sources of entangled qubits, as illustrated in Fig.~\ref{fig_scenarioq}. Following~\cite{biloc1_PRL}, we call {\it bilocal} the correlations that can be described as resulting from measurements of two independent hidden states $\lambda_1$ and $\lambda_2$, each produced by one of the two sources. Conversely, a correlation that cannot be described in such a way is called {\it non-bilocal}.

We first introduce more formally the concept of bilocality in Section~\ref{sec_characterization}, and show how the bilocality assumption can be tested. A first approach, developped in subsection~\ref{subsec_explicit_models}, is to look for explicit bilocal decompositions; we introduce efficient representations for (bi-)local models, which help the search for explicit decompositions, and where the bilocality assumption takes a very simple form.
Another approach is to test Bell-like inequalities, which are satisfied by bilocal correlations but can be violated by non-bilocal correlations; in subsection~\ref{subsec_nonlinear_ineq} we derive such (nonlinear) {\it bilocal inequalities} (eqs.~(\ref{ineq_22}),~(\ref{ineq_14}) and~(\ref{ineq_13})), which typically take the form
\ba
\sqrt{|I|} + \sqrt{|J|} \ \leq \ 1 \,,
\ea
where $I$ and $J$ are linear combinations of probabilities $P(a,b,c|x,y,z)$ (see eqs.~(\ref{def_I_22}--\ref{def_J_22}),~(\ref{def_I_14}--\ref{def_J_14}) and~(\ref{def_I_13}--\ref{def_J_13})).

Section~\ref{sec_qviol} is devoted to the study of how the correlations produced in quantum entanglement swapping experiments violate our bilocal inequalities. We analyse the situations where Bob performs a complete Bell state measurement in subsection~\ref{subsec_PQ_14}, and where he performs partial Bell state measurements that allows him to distinguish different pairs of Bell states (subsection~\ref{subsubsec_PQ_22}), or to distinguish two Bell states, while the other two give the same outcome (subsection~\ref{subsubsec_PQ_13}). In all cases we study the resistance to white noise of the quantum violations, and find that the required visibilities for demonstrating non-bilocality are significantly lower than for demonstrating Bell nonlocality. The resistance to detection inefficiencies is also analysed in some simple cases for the complete Bell state measurement in subsection~\ref{subsec_det_loop}.

In Section~\ref{sec_misc}, we address further issues on quantum non-bilocality. We study a trade-off between the resistance to noise of nonlocality and non-bilocality for quantum correlations (subsection~\ref{subsec_pareto}), and show that the two are not necessarily correlated but that the maximization of one is made at the expense of the other. We then investigate possible violations of the bilocality assumption using non-maximally entangled states (subsection~\ref{subsec_nonmaxent}). We also address the question of classically simulating noisy entanglement swapping correlations, and introduce two protocols (with and without communication) for that in subsection~\ref{subsec_simul}.

Finally, in Section~\ref{sec_triloc} we come back to our justification for the assumption of independent sources, and to the idea that it is actually already implicitly used in standard Bell experiments. We illustrate this claim by showing that the assumption of local causality with independent sources in a bipartite Bell test is equivalent to an assumption of {\it trilocality} in a four-partite experiment.

\section{Characterizing bilocal correlations}
\label{sec_characterization}

\subsection{The bilocality assumption}

We consider the scenario depicted in Figure~\ref{fig_scenariol}, with three parties sharing two sources of independent hidden states; this is the simplest case where our assumption of independent sources of hidden states makes sense.

In such a tripartite scenario, Bell's locality assumption~\cite{bell} reads (as we recalled in eq.~(\ref{eq_locality}))
\ba
P(a,\!b,\!c|x,\!y,\!z) = \! \int \!\! \mathrm{d} \lambda \, \rho(\lambda) \, P(a|x,\!\lambda)P(b|y,\!\lambda)P(c|z,\!\lambda) . \quad \ \label{eq_def_locality1}
\ea
Here given the state $\lambda$, the outputs $a$, $b$ and $c$ of the three parties, for inputs $x$, $y$ and $z$, are determined respectively by the local distributions $P(a|x,\lambda)$, $P(b|y,\lambda)$ and $P(c|z,\lambda)$. The hidden states $\lambda$ follow the distribution $\rho(\lambda)$, normalized such that $\int \! \mathrm{d} \lambda \, \rho(\lambda) = 1$.

If we now assume that the response of the three parties depend only on the states $\lambda_1$ or $\lambda_2$ characterizing the systems that they receive from the sources $S_1$ or $S_2$, respectively, we write:
\ba
P(a,\!b,\!c|x,\!y,\!z) &\!=\!& \int\!\!\!\!\!\int \! \mathrm{d} \lambda_1 \mathrm{d} \lambda_2 \, \rho(\lambda_1,\lambda_2) \nonumber \\
&& \quad \ P(a|x,\!\lambda_1)P(b|y,\!\lambda_1,\!\lambda_2)P(c|z,\!\lambda_2) , \quad \ \ \label{eq_def_locality2}
\ea
Note that without making any further assumption, equation~(\ref{eq_def_locality2}) is equivalent to~(\ref{eq_def_locality1}); indeed, $\rho(\lambda_1,\lambda_2)$ could be different from zero only when $\lambda_1=\lambda_2=\lambda$ to recover~(\ref{eq_def_locality1}).

Now we introduce our crucial assumption: since the two quantum sources $S_1$ and $S_2$ are supposed to be independent, we assume that this property carries over to the local model, and therefore the distribution of the hidden states $\lambda_1$ and $\lambda_2$ should factorize, in the form
\ba
 \rho(\lambda_1,\lambda_2) = \rho_1(\lambda_1) \, \rho_2(\lambda_2) \,. \label{eq_constr_biloc_rho}
\ea
Together with~(\ref{eq_def_locality2}), this defines our assumption of {\it bilocality}, as already expressed in equation~(\ref{eq_bilocality}). The hidden states $\lambda_1$ and $\lambda_2$ now follow independent distributions $\rho_1(\lambda_1)$ and $\rho_2(\lambda_2)$, such that $\int \! \mathrm{d} \lambda_1 \, \rho_1(\lambda_1) = \int \! \mathrm{d} \lambda_2 \, \rho_2(\lambda_2) = 1$.

Note that, as in the standard case of Bell locality, no restrictions are made on the sets on which $\lambda_1$, $\lambda_2$ are distributed (apart from the fact that they must be measurable). Expanding the results of~\cite{biloc1_PRL}, we show in the subsections~\ref{subsec_explicit_models} and~\ref{subsec_nonlinear_ineq} below that for finite numbers of inputs and outputs, eqs.~(\ref{eq_def_locality2}) and~(\ref{eq_constr_biloc_rho}) lead nonetheless to implementable tests of bilocality, by looking for explicit bilocal decompositions or by testing (non-linear) Bell-like inequalities. Before that, let us briefly mention some general properties of the set of bilocal correlations.

\subsection{Topology of the bilocal set}

Any bilocal correlation is by construction also local; the set of bilocal correlations (``bilocal set'', $\mathcal{B}$) is therefore included in the set of local correlations (``local set'', $\mathcal{L}$): $\mathcal{B} \subseteq \mathcal{L}$.

It is well known, and clear from the definition~(\ref{eq_def_locality1}), that the local set is convex. On the other hand, because of the nonlinear constraint~(\ref{eq_constr_biloc_rho}), a mixture of bilocal correlations is not necessarily bilocal: the bilocal set is not convex. Deterministic local correlations are bilocal; since they are the extremal points of the local set, $\mathcal{L}$ is actually the convex hull of $\mathcal{B}$.

One can further show (see Appendix~\ref{app_topology}) that the bilocal set is connected, and that its restriction to subspaces where the marginal probability distribution of Alice (or Charlie) is fixed is star-convex; star-convexity does not however hold for the whole bilocal set.

\subsection{Explicit bilocal decompositions}

\label{subsec_explicit_models}

From now on we consider scenarios with finite numbers of possible inputs and outputs.

In that case, the local set ${\cal L}$ forms a convex polytope~\cite{pitowsky}. The description of the local polytope as the convex hull of a finite set of extremal points---corresponding to deterministic local correlations---allows one to use efficient numerical approaches based on linear programming to determine if a correlation $P$ is local.
Alternatively, the local polytope can be described in terms of its facets---corresponding to (possibly trivial) Bell inequalities---which can be enumerated algorithmically for a small enough number of inputs and outputs.
Hence, in order to determine whether a correlation $P$ is local or not, one can either solve a linear programming problem, or check that all Bell inequalities are satisfied.

Because the bilocal set is not convex, one cannot use standard Bell inequalities to distinguish the bilocal and the non-bilocal correlations. The bilocal set is significantly more difficult to characterize than the local set.
Still, similar approaches can be used: one can describe the question whether a given correlation $P$ is bilocal or not as a non-convex feasibility problem, or alternatively (as we will see in the next subsection~\ref{subsec_nonlinear_ineq}), one can derive non-linear inequalities that are satisfied by any bilocal point. However, we do not have a systematic practical approach to obtain such ``bilocal inequalities''; and even in the simplest case we consider, we do not have a complete set of inequalities that would be sufficient to define the set $\mathcal{B}$.

We start here by giving alternative formulations for the bilocality assumption~(\ref{eq_def_locality2}--\ref{eq_constr_biloc_rho}), which will prove more handy to use for practical purposes, in particular when looking for explicit bilocal decompositions.

\subsubsection{Bilocal decompositions onto deterministic correlations, \\ with weights $q_{\bar{\alpha}\bar{\beta}\bar{\gamma}}$}

Consider a local correlation $P$, written in the form~(\ref{eq_def_locality2}). It is well known~\cite{fine} that Alice's local response function $P(a|x,\lambda_1)$ can (without any loss of generality) be taken to be deterministic, i.e., such that it assigns a unique measurement output $a$ to every input $x$: any randomness used locally by Alice can indeed always be thought of as being included in the shared random variable $\lambda_1$. For a finite number of possible measurement inputs and outputs, there is a finite number of such deterministic strategies corresponding to an assignment of an output $\alpha_x$ to each of Alice's $N$ possible inputs $x$. We  label each of these strategies with the string $\bar{\alpha}=\alpha_1\ldots \alpha_N$ and denote the corresponding response function $P_{\bar{\alpha}}(a|x) = \delta_{a,\alpha_x}$ (with $\delta_{m,n} = 1$ if $m=n$, $\delta_{m,n} = 0$ otherwise). Similarly, the response functions $P(b|y,\lambda_1,\lambda_2)$ and $P(c|z,\lambda_2)$ can also be taken deterministic; we label the associated strategies $\bar{\beta}$ and $\bar{\gamma}$ and the corresponding response functions are $P_{\bar{\beta}}(b|y) = \delta_{b,\beta_y}$ and $P_{\bar{\gamma}}(c|z) = \delta_{c,\gamma_z}$.

Integrating over the set $\Lambda^{12}_{\bar{\alpha}\bar{\beta}\bar{\gamma}}$ of all pairs $(\lambda_1,\lambda_2)$ that specify the strategies $\bar{\alpha}$, $\bar{\beta}$, and $\bar{\gamma}$ for Alice, Bob and Charlie resp., we can write~(\ref{eq_def_locality2}) as
\ba
P(a,\!b,\!c|x,\!y,\!z) = \sum_{\bar{\alpha},\bar{\beta},\bar{\gamma}} q_{\bar{\alpha}\bar{\beta}\bar{\gamma}} \, P_{\bar{\alpha}}(a|x)P_{\bar{\beta}}(b|y)P_{\bar{\gamma}}(c|z) \quad
\label{eq_locality_sum}
\ea
with $q_{\bar{\alpha}\bar{\beta}\bar{\gamma}}=\int\!\!\!\int_{\Lambda^{12}_{\bar{\alpha}\bar{\beta}\bar{\gamma}}} \mathrm{d} \lambda_1 \mathrm{d} \lambda_2 \, \rho(\lambda_1,\lambda_2) \geq 0$ and $\sum_{\bar{\alpha}\bar{\beta}\bar{\gamma}} q_{\bar{\alpha}\bar{\beta}\bar{\gamma}}=1$. Eq.~(\ref{eq_locality_sum}) corresponds to the well-known decomposition of local correlations as a convex sum of deterministic strategies, where the weights $q_{\bar{\alpha}\bar{\beta}\bar{\gamma}}$ can be understood as the probabilities assigned by the source to the strategies $\bar{\alpha}$, $\bar{\beta}$ and $\bar{\gamma}$.

Since $\bar{\alpha}$ is specified here by $\lambda_1$ and $\bar{\gamma}$ is specified by $\lambda_2$, then (with obvious notations) $\cup_{\bar{\beta}} \Lambda^{12}_{\bar{\alpha}\bar{\beta}\bar{\gamma}} = \Lambda^{12}_{\bar{\alpha}\bar{\gamma}} = \Lambda^{1}_{\bar{\alpha}} \times \Lambda^{2}_{\bar{\gamma}}$, and
\ba
&q_{\bar{\alpha}\bar{\gamma}} = \sum_{\bar{\beta}} q_{\bar{\alpha}\bar{\beta}\bar{\gamma}} = \int\!\!\!\int_{\Lambda^{1}_{\bar{\alpha}} \times \Lambda^{2}_{\bar{\gamma}}} \mathrm{d} \lambda_1 \mathrm{d} \lambda_2 \, \rho(\lambda_1,\lambda_2), \label{q_ac} \\
&q_{\bar{\alpha}} = \sum_{\bar{\gamma}} q_{\bar{\alpha}\bar{\gamma}} = \int\!\!\!\int_{\Lambda^{1}_{\bar{\alpha}} \times \Lambda^{2}} \mathrm{d} \lambda_1 \mathrm{d} \lambda_2 \, \rho(\lambda_1,\lambda_2), \label{q_a} \\
&q_{\bar{\gamma}} = \sum_{\bar{\alpha}} q_{\bar{\alpha}\bar{\gamma}} = \int\!\!\!\int_{\Lambda^{1} \times \Lambda^{2}_{\bar{\gamma}}} \mathrm{d} \lambda_1 \mathrm{d} \lambda_2 \, \rho(\lambda_1,\lambda_2), \label{q_c}
\ea
where $\Lambda^{1} = \cup_{\bar{\alpha}} \Lambda^{1}_{\bar{\alpha}}$ and $\Lambda^{2} = \cup_{\bar{\gamma}} \Lambda^{2}_{\bar{\gamma}}$ are the state spaces of the variables $\lambda_1$ and $\lambda_2$.

Let us now assume that $P$ is bilocal. One can see from~(\ref{q_ac}--\ref{q_c}) that the independence condition~(\ref{eq_constr_biloc_rho}) implies that
\ba
{\mathrm{for \ all}} \ \bar{\alpha},\bar{\gamma}, \quad q_{\bar{\alpha}\bar{\gamma}} \ = \ q_{\bar{\alpha}} \, q_{\bar{\gamma}} \,. \label{eq_constr_biloc_qabc}
\ea
The interpretation is clear: the strategies $\bar{\alpha}$ and $\bar{\gamma}$ being determined by two independent sources, their probabilities should be independent.

Conversely, any correlation $P(a,\!b,\!c|x,\!y,\!z)$ satisfying~(\ref{eq_locality_sum}) and~(\ref{eq_constr_biloc_qabc}) can be written in the form~(\ref{eq_def_locality2}). Indeed, since $q_{\bar{\alpha}\bar{\gamma}}=q_{\bar{\alpha}} q_{\bar{\gamma}}$, we can write $q_{\bar{\alpha}\bar{\beta}\bar{\gamma}}=q_{\bar{\alpha}} q_{\bar{\gamma}} q_{\bar{\beta}|\bar{\alpha}\bar{\gamma}}$. Inserting this expression in~(\ref{eq_locality_sum}) and defining $P_{\bar{\alpha},\bar{\gamma}}(b|y)=\sum_{\bar{\beta}} q_{\bar{\beta}|\bar{\alpha}\bar{\gamma}} P_{\bar{\beta}}(b|y)$, we then find that $P(a,\!b,\!c|x,\!y,\!z) = \sum_{\bar{\alpha},\bar{\gamma}} q_{\bar{\alpha}} q_{\bar{\gamma}} \, P_{\bar{\alpha}}(a|x)P_{\bar{\alpha}\bar{\gamma}}(b|y)P_{\bar{\gamma}}(c|z)$, which is clearly of the form~(\ref{eq_def_locality2}).
We thus conclude that a tripartite correlation is bilocal if and only if it admits the decomposition~(\ref{eq_locality_sum}) with the restriction~(\ref{eq_constr_biloc_qabc}).

Such a description of bilocal correlations is easier to deal with than the defining assumptions~(\ref{eq_def_locality2}) and~(\ref{eq_constr_biloc_rho}), as it involves only a finite number of coefficients $q_{\bar{\alpha}\bar{\beta}\bar{\gamma}}$. One can thus now determine whether a correlation $P$ is bilocal by searching such weights $q_{\bar{\alpha}\bar{\beta}\bar{\gamma}} \geq 0$, with the linear constraints that they must reproduce the correlation $P$ as in~(\ref{eq_locality_sum}), and the quadratic constraints of bilocality~(\ref{eq_constr_biloc_qabc}).

An even more compact representation of bilocal decompositions can however be given in the following way.

\subsubsection{Decompositions in terms of ``correlators'' $e_{\bar{i}\bar{j}\bar{k}}$}

\label{subsection_correlators}

For a given local (or bilocal) correlation $P$, the decomposition (\ref{eq_locality_sum}) is in general not unique; Eq. (\ref{eq_locality_sum}) only imposes a limited number of constraints on the weights $q_{\bar{\alpha}\bar{\beta}\bar{\gamma}}$, which are not enough to fix all of them.
When dealing with a local decomposition, it is convenient to use a parametrization that clearly separates the parameters that are fixed, and those that are internal degrees of freedom of the local model. A nice way to do it is to transform the weights $q_{\bar{\alpha}\bar{\beta}\bar{\gamma}}$ into ``correlators'' $e_{\bar{i}\bar{j}\bar{k}}$.

Since we will use this approach in the context of bilocality, we present it here in the 3-partite case; note however that this representation can be useful in more general studies of \mbox{(non-)locality}---not only of bilocality---and it can easily be generalised to any $N$-partite case. Also, for simplicity we consider here a scenario with binary inputs and outputs; the generalization to other scenarios can be cumbersome but is rather straightforward (see subsection~\ref{subsubsec_diff_22_14} and Appendix~\ref{app_explicit_decomps} for instance for the case where Bob has only 1 possible input, and 4 or 3 outputs).

\medskip

For binary inputs and outputs, Alice, Bob and Charlie's strategies $\bar{\alpha},\bar{\beta},\bar{\gamma}$ simply contain two bits: $\bar{\alpha} = \alpha_0\alpha_1$, etc. Let us then define, for a particular local decomposition of $P$ in terms of weights $q_{\bar{\alpha}\bar{\beta}\bar{\gamma}}$, and for $\bar{i}=i_0i_1,\bar{j}=j_0j_1$ and $\bar{k}=k_0k_1 \in \{00,01,10,11\}$, the coefficients
\ba
e_{\bar{i}\bar{j}\bar{k}} = \sum_{\bar{\alpha}\bar{\beta}\bar{\gamma}} (-1)^{\bar{\alpha}\cdot\bar{i}+\bar{\beta}\cdot\bar{j}+\bar{\gamma}\cdot\bar{k}} q_{\bar{\alpha}\bar{\beta}\bar{\gamma}} \, , \label{def_eijk}
\ea
where $\bar{\alpha}\cdot\bar{i} = \alpha_0 i_0 + \alpha_1 i_1$, etc. The set of coefficients $\{e_{\bar{i}\bar{j}\bar{k}}\}$ is equivalent to the set of weights $\{q_{\bar{\alpha}\bar{\beta}\bar{\gamma}}\}$ (eq.~(\ref{def_eijk}) is actually a discrete Fourier transformation), and~(\ref{def_eijk}) can easily be inverted to obtain
\ba
q_{\bar{\alpha}\bar{\beta}\bar{\gamma}} = 2^{-6} \sum_{\bar{i}\bar{j}\bar{k}} (-1)^{\bar{\alpha}\cdot\bar{i}+\bar{\beta}\cdot\bar{j}+\bar{\gamma}\cdot\bar{k}} e_{\bar{i}\bar{j}\bar{k}} \, .
\ea
Thus, both representations (in terms of coefficients $q_{\bar{\alpha}\bar{\beta}\bar{\gamma}}$ or $e_{\bar{i}\bar{j}\bar{k}}$) can be used to unambiguously define the local decomposition.

We call the coefficients $e_{\bar{i}\bar{j}\bar{k}}$ ``correlators'', for the following reason: defining\footnote{When expressing correlations, it is often more convenient to consider $\pm 1$-valued outputs. Throughout the paper, we'll use lowercase variables for bit values $0/1$, and uppercase variables for the corresponding bit values $\pm1$, as for instance in ${\mathrm A}_x = (-1)^{\alpha_x}$.}, for a strategy $\bar{\alpha}\bar{\beta}\bar{\gamma}$, ${\mathrm A}_x = (-1)^{\alpha_x}$, ${\mathrm B}_y = (-1)^{\beta_y}$, and ${\mathrm C}_z = (-1)^{\gamma_z}$, we have
\ba
e_{\bar{i}\bar{j}\bar{k}} = \moy{{\mathrm A}_0^{i_0}{\mathrm A}_1^{i_1} {\mathrm B}_0^{j_0}{\mathrm B}_1^{j_1} {\mathrm C}_0^{k_0}{\mathrm C}_1^{k_1}}_{\{q_{\bar{\alpha}\bar{\beta}\bar{\gamma}}\}} \, . \label{eq_eijk_moy}
\ea
where the average value is computed with the weights $q_{\bar{\alpha}\bar{\beta}\bar{\gamma}}$ 
(note that for a deterministic strategy, all values ${\mathrm A}_x$, ${\mathrm B}_y$ and ${\mathrm C}_z$ are defined simultaneously).

Now, when $\bar{i} \neq \bar{1}$, $\bar{j} \neq \bar{1}$ and $\bar{k} \neq \bar{1}$ (with the notation $\bar{1} = 11$), at most one term ${\mathrm A}_x$, ${\mathrm B}_y$ and ${\mathrm C}_z$ per party appears non-trivially in the average above. Hence, this average value can be obtained directly from the correlation $P$, and the corresponding correlators\footnote{More generally, for more than two inputs per party, the fixed correlators are those where the corresponding product averaged in~(\ref{eq_eijk_moy}) only contains terms that refer to at most one input per party.} $e_{\bar{i}\bar{j}\bar{k}}$ are therefore {\it fixed by $P$}:
\ba
{\mathrm{for}} \ \bar{i} \neq \bar{1}, \bar{j} \neq \bar{1}, \ {\mathrm{and}} \ \bar{k} \neq \bar{1}, \hspace{1.5cm} \nonumber \\
e_{\bar{i}\bar{j}\bar{k}} = \moy{A_0^{i_0}A_1^{i_1} B_0^{j_0}B_1^{j_1} C_0^{k_0}C_1^{k_1}}_P \label{eijk_eq_constr}
\ea
with now $A_x = (-1)^{a_x}$, where $a_x \in \{0,1\}$ is Alice's output for the input $x$, and similarly for the other two parties. The average is now computed from the correlation $P$. For instance, one gets $e_{10,10,10} = \moy{A_0 B_0 C_0}_P$, and $e_{01,00,00} = \moy{A_1}_P$. Note that in particular, with $\bar{0} = 00$, one has $e_{\bar{0}\bar{0}\bar{0}} = 1$ by normalization.

On the other hand, when $\bar{i} = \bar{1}$, $\bar{j} = \bar{1}$ or $\bar{k} = \bar{1}$, the average value in (\ref{eq_eijk_moy}) cannot be obtained from $P$, since the measurement results $A_0$ and $A_1$, $B_0$ and $B_1$, $C_0$ and $C_1$ are incompatible; the corresponding correlators are {\it internal degrees of freedom} of the local model, only constrained by the nonnegativity of $q_{\bar{\alpha}\bar{\beta}\bar{\gamma}}$, i.e.,
\ba
{\mathrm{for \ all}} \ \bar{\alpha},\bar{\beta},\bar{\gamma}, \quad \sum_{\bar{i}\bar{j}\bar{k}} \ (-1)^{\bar{\alpha}\cdot\bar{i}+\bar{\beta}\cdot\bar{j}+\bar{\gamma}\cdot\bar{k}} \ e_{\bar{i}\bar{j}\bar{k}} \geq 0 \, . \label{eq_constr_pos_eijk}
\ea

\medskip

Coming back to the bilocality constraint: in the correlators representation, one can easily check that the condition (\ref{eq_constr_biloc_qabc}) translates into the following constraints:
\ba
{\mathrm{for \ all}} \ \bar{i},\bar{k}, \quad e_{\bar{i}\bar{0}\bar{k}} \ = \ e_{\bar{i}\bar{0}\bar{0}} \ e_{\bar{0}\bar{0}\bar{k}} \,. \label{eq_constr_biloc_eijk}
\ea
Now, when $\bar{i} \neq \bar{1}$ and $\bar{k} \neq \bar{1}$, the correlators that appear in~(\ref{eq_constr_biloc_eijk}) are already fixed according to~(\ref{eijk_eq_constr}), and the constraint is indeed satisfied, as a consequence of the fact that for bilocal correlations, $P(a,\!c|x,\!z) = P(a|x) P(c|z)$\footnote{The fact that $P(a,\!c|x,\!z) = P(a|x) P(c|z)$ directly follows from the bilocality condition~(\ref{eq_bilocality}): after summing over Bob's outputs $b$, the two integrals over $\lambda_1$ and $\lambda_2$ factorize. Note that this equality also holds for quantum correlations established from independent sources, as in Fig.~\ref{fig_scenarioq}.} (or $\moy{A_x C_z} = \moy{A_x} \moy{C_z}$). When $\bar{i} = \bar{1}$ and $\bar{k} \neq \bar{1}$, or vice versa, (\ref{eq_constr_biloc_eijk}) gives linear constraints on the free correlators $e_{\bar{1}\bar{0}\bar{k}}$ and $e_{\bar{i}\bar{0}\bar{1}}$. There finally remains only one\footnote{For more than 2 inputs for Alice and Charlie, there will remain more than one quadratic constraint, but the correlators representation will still significantly simplify the search for explicit bilocal decompositions.} quadratic constraint in the case where $\bar{i}  = \bar{k} = \bar{1}$: $e_{\bar{1}\bar{0}\bar{1}} = e_{\bar{1}\bar{0}\bar{0}} \, e_{\bar{0}\bar{0}\bar{1}}$.

Because the representation in terms of correlators nicely separates the fixed and free parameters of the (bi)local decomposition, it simplifies the search for explicit bilocal decompositions quite significantly. We now present how this problem can be tackled in practice.

\subsubsection{Looking for explicit bilocal decompositions}

\label{subsubsec_numerical_search}

To determine whether a given correlation $P$ is bilocal or not, one can now look whether there exist correlators $e_{\bar{i}\bar{j}\bar{k}}$---some of which are fixed by~(\ref{eijk_eq_constr})---that satisfy the nonnegativity constraint~(\ref{eq_constr_pos_eijk}) and the bilocality assumption~(\ref{eq_constr_biloc_eijk}).

This non-convex feasibility problem can be addressed from two different perspectives. First, one can try a heuristic search that will provide an explicit bilocal model if successful, thus proving $P\in\mathcal{B}$---unless an exhaustive search can be undergone, a negative result will however be inconclusive. Secondly, one can try to solve a convex relaxation~\cite{branchnbound1966} of the quadratic constraints~(\ref{eq_constr_biloc_eijk}) in the non-convex problem---a negative result proving $P\notin\mathcal{B}$.
Both approaches lead naturally to the usage of numerical algorithms to solve the non-convex problem or its convex relaxations\footnote{In our numerical tests, we used the standard optimization toolbox of MATLAB, which implements the algorithms described in~\cite{waltz2006interior}, and the BMIBNB solver of YALMIP for convex relaxations.}.

In practice, one may be interested in the robustness of a non-bilocal correlation to experimental imperfections, such as noise (see in particular subsections~\ref{subsubsec_PQV_14} and~\ref{subsec_partialBSM}) or detection inefficiencies (see subsection~\ref{subsec_det_loop}). One can then transform the above feasibility problem into an optimization problem; for instance, one may want to maximize the noise that is tolerated by a correlation (or minimize the visibility of a noisy correlation of the form $P(V) = V P + (1-V) P_0$, see eq.~(\ref{eq_PQV}) below) before it becomes bilocal. The brute force search will give an upper bound on the tolerated noise (a lower bound on the visibility threshold $V_{biloc}$, as defined in subsection~\ref{subsubsec_PQV_14} below), while the convex relaxation will give a lower bound on the tolerated noise (an upper bound on $V_{biloc}$).

Examples of explicit bilocal decompositions, for the quantum correlations studied in section~\ref{sec_qviol}, are given in Appendix~\ref{app_explicit_decomps}.

\subsection{Non-linear inequalities for bilocal correlations}

\label{subsec_nonlinear_ineq}

As mentioned earlier, another approach to study the \mbox{(non-)bilocality} of given correlations is to test Bell-type inequalities. Indeed, one can derive in some cases analytical constraints satisfied by all bilocal correlations; if a correlation is found to violate these constraints, then this implies that the correlation is non-bilocal. These non-linear \emph{bilocal inequalities} thus provide convenient tests of non-bilocality, which are directly implementable in experimental demonstrations.

We start by deriving such a bilocal inequality for the case where the three parties all have binary inputs and outputs. We then show that our first inequality implies similar inequalities for the case where Bob now has only one possible input, with 4 or 3 possible outputs (while Alice and Charlie still have binary inputs and outputs). These cases will be relevant for entanglement swapping experiments, as we will later see in section~\ref{sec_qviol}.

\subsubsection{A bilocal inequality for binary inputs and outputs}

We thus consider first the scenario where Alice, Bob and Charlie have binary inputs and outputs $x,y,z,a,b,c\in\{0,1\}$.

Let us define, for a given correlation $P^{22}$, the tripartite correlation terms
\ba
\moy{A_x B_y C_z}_{P^{22}} &=& \sum_{a,b,c} \, (-1)^{a + b + c} \, P^{22} ( a, b, c | x, y, z ) \nonumber
\ea
and the following linear combinations $I^{22}, J^{22}$:
\ba
I^{22} &=& \frac{1}{4} \sum_{x,z=0,1} \ \moy{A_x B_0 C_z}_{P^{22}} \, , \label{def_I_22} \\
J^{22} &=& \frac{1}{4} \sum_{x,z=0,1} \ (-1)^{x+z} \moy{A_x B_1 C_z}_{P^{22}} \, . \label{def_J_22}
\ea

As we show below, if $P^{22}$ is bilocal, then the following nonlinear inequality necessarily holds:
\begin{equation}
 \sqrt{|I^{22}|} + \sqrt{|J^{22}|} \ \leq \ 1 \,. \label{ineq_22}
\end{equation}

\medskip

\begin{proof}

By assumption, $P^{22}$ has a bilocal decomposition of the form~(\ref{eq_bilocality}). Defining $\moy{A_x}_{\lambda_1} = \sum_a (-1)^a P^{22}(a|x,\lambda_1)$, and with similar definitions for $\moy{B_y}_{\lambda_1,\lambda_2}$ and $\moy{C_z}_{\lambda_2}$, one gets
\ba
I^{22} & \!= & \frac{1}{4} \! \int\!\!\!\!\!\int \! \mathrm{d} \lambda_1 \mathrm{d} \lambda_2 \, \rho_1(\lambda_1) \, \rho_2(\lambda_2) \nonumber \\
&& \quad \ \big(\moy{A_0}_{\lambda_1} \!\! + \! \moy{A_1}_{\lambda_1} \big) \moy{B_0}_{\lambda_1,\lambda_2} \big(\moy{C_0}_{\lambda_2} \!\! + \! \moy{C_1}_{\lambda_2} \big) \,. \nonumber
\ea
Using the fact that $|\moy{B_0}_{\lambda_1,\lambda_2}| \leq 1$,
\ba
&& \hspace{-.1cm} |I^{22}| \nonumber \\
&& \hspace{-.1cm} \leq \frac{1}{4} \! \int\!\!\!\!\!\int \! \mathrm{d} \lambda_1 \mathrm{d} \lambda_2 \rho_1(\lambda_1) \rho_2(\lambda_2) \big|\moy{A_0}_{\!\lambda_1} \!\! + \! \moy{A_1}_{\!\lambda_1} \big| \big|\moy{C_0}_{\!\lambda_2} \!\! + \! \moy{C_1}_{\!\lambda_2} \big| \nonumber \\
&& \hspace{-.1cm} \leq \!\!\int\!\! \mathrm{d} \lambda_1 \rho_1(\lambda_1) \frac{\big|\moy{A_0}_{\!\lambda_1} \!\! + \! \moy{A_1}_{\!\lambda_1} \!\big|}{2} \!\!\times\!\! \!\int\!\! \mathrm{d} \lambda_2 \rho_2(\lambda_2) \frac{\big|\moy{C_0}_{\!\lambda_2} \!\! + \! \moy{C_1}_{\!\lambda_2} \!\big|}{2}, \nonumber
\ea
and one can show similarly, that
\ba
&& \hspace{-.1cm} |J^{22}| \nonumber \\
&& \hspace{-.1cm} \leq \!\!\int\!\! \mathrm{d} \lambda_1 \rho_1(\lambda_1) \frac{\big|\moy{A_0}_{\!\lambda_1} \!\! - \! \moy{A_1}_{\!\lambda_1} \!\big|}{2} \!\!\times\!\! \!\int\!\! \mathrm{d} \lambda_2 \rho_2(\lambda_2) \frac{\big|\moy{C_0}_{\!\lambda_2} \!\! - \! \moy{C_1}_{\!\lambda_2} \!\big|}{2} . \nonumber
\ea
Now, for any $r, s, r', s' \geq 0$, the inequality $\sqrt{rs} + \sqrt{r' s'} \leq \sqrt{r + r'} \sqrt{s + s'}$ holds. Applied to the above two bounds on $|I^{22}|$ and $|J^{22}|$, we obtain
\ba
&& \hspace{-.1cm} \sqrt{|I^{22}|} + \sqrt{|J^{22}|} \nonumber \\
&& \hspace{-.1cm} \leq \sqrt{\!\!\int\!\! \mathrm{d} \lambda_1 \rho_1(\lambda_1) \! \left( \frac{\big|\moy{A_0}_{\!\lambda_1} \!\! + \! \moy{A_1}_{\!\lambda_1} \!\big|}{2} + \frac{\big|\moy{A_0}_{\!\lambda_1} \!\! - \! \moy{A_1}_{\!\lambda_1} \!\big|}{2} \right)} \nonumber \\[-2mm]
&& \quad \times \sqrt{\!\!\int\!\! \mathrm{d} \lambda_2 \rho_2(\lambda_2) \! \left( \frac{\big|\moy{C_0}_{\!\lambda_2} \!\! + \! \moy{C_1}_{\!\lambda_2} \!\big|}{2} + \frac{\big|\moy{C_0}_{\!\lambda_2} \!\! - \! \moy{C_1}_{\!\lambda_2} \!\big|}{2} \right)} \,. \nonumber
\ea
Furthermore, $| \moy{A_0}_{\lambda_1} + \moy{A_1}_{\lambda_1} |/2 + | \moy{A_0}_{\lambda_1} - \moy{A_1}_{\lambda_1} |/2 = \max (|\moy{A_0}_{\lambda_1}|, |\moy{A_1}_{\lambda_1}|) \leq 1$ and similarly, $| \moy{C_0}_{\lambda_2} + \moy{C_1}_{\lambda_2} |/2 + | \moy{C_0}_{\lambda_2} - \moy{C_1}_{\lambda_2} |/2 \leq 1$. After integrating over $\lambda_1$ and $\lambda_2$ in the previous expressions, we obtain inequality~(\ref{ineq_22}).
\end{proof}

\begin{figure}
\begin{center}
\epsfxsize=7.2cm
\epsfbox{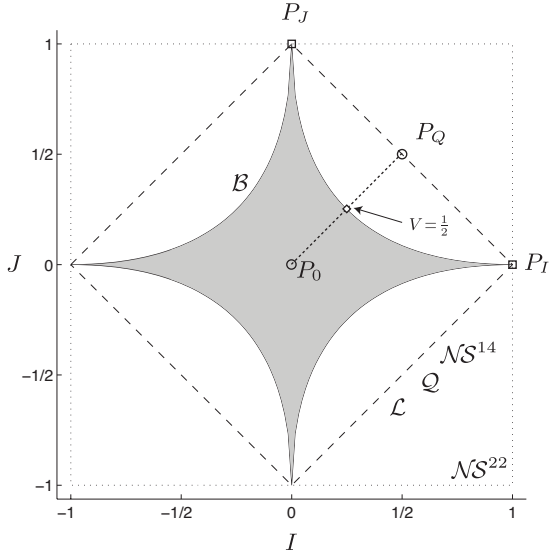}
\caption{Projection of the tripartite correlation space in the $(I,J)$ plane, for $I = I^{22}$ or $I^{14}$ and $J = J^{22}$ or $J^{14}$, as defined in~(\ref{def_I_22}--\ref{def_J_22}) or~(\ref{def_I_14}--\ref{def_J_14}), for the \mbox{``22-''} and \mbox{``14-''cases}, resp. The non-convex bilocal set ${\cal B}$ is delimited by the four portions of parabolas, corresponding to the inequality $\sqrt{|I|}+\sqrt{|J|} \leq 1$ (eqs.~(\ref{ineq_22}) or~(\ref{ineq_14})). It is included in the local set ${\cal L}$ delimited by the dashed lines, for which $|I|+|J| \leq 1$. The set ${\cal Q}$ of quantum correlations is also limited in this plane by $|I|+|J| \leq 1$. In the \mbox{22-case}, the non-signaling polytope ${\cal N\!S}^{22}$ is delimited by the outer dotted square, defined by $\max(|I^{22}|,|J^{22}|) \leq 1$; in the \mbox{14-case}, the projection of the non-signaling polytope ${\cal N\!S}^{14}$ coincides with that of the local and quantum sets.
\\
The figure can also be understood as a 2-dimensional slice of the correlation space, containing the quantum correlation $P_Q = P_Q^{22}$~(\ref{eq_PQ_22}) or $P_Q^{14}$~(\ref{eq_PQ_14}), the fully random correlation $P_0 = P_0^{22}$ or $P_0^{14}$, and the bilocal correlations $P_I=P_I^{22}$ or $P_I^{14}$ and $P_J=P_J^{22}$ or $P_J^{14}$, as defined in footnotes~\ref{footnote_local_decomp_PQ14} and~\ref{footnote_local_decomp_PQ22}. In this slice, the bilocal set is star-convex. One can see that the quantum correlation $P_Q = \demi P_I + \demi P_J$ is local, but not bilocal. When adding some white noise, it enters the bilocal set for visibilities $V \leq \demi$ (see subsection~\ref{subsubsec_PQV_14}).}
\label{fig_IJ_plane_22_14}
\end{center}
\end{figure}

A projection of the correlation space onto the $(I^{22},J^{22})$ plane is shown on Figure~\ref{fig_IJ_plane_22_14}. Note that the bilocal  inequality~(\ref{ineq_22}) is tight in this plane, i.e., any values of $I^{22}$ and $J^{22}$ such that $\sqrt{|I^{22}|} + \sqrt{|J^{22}|} \leq 1$ can be obtained by a bilocal correlation; see Table~\ref{table_IJ22} in Appendix~\ref{app_explicit_decomps} for an explicit bilocal decomposition.
One can show on the other hand that the local correlations satisfy $|I^{22}|+|J^{22}| \leq 1$ (recall that $\mathcal{L}$ is the convex hull of $\mathcal{B}$), which can be understood as Bell inequalities\footnote{For instance, the inequality $I^{22} + J^{22} \leq 1$ can be written as (a quarter of) the sum of two equivalent facet inequalities of the tripartite local polytope (it is therefore not a facet itself): $\moy{A_0B_0C_0+A_1B_1C_1+A_1B_0C_0-A_0B_1C_1} \leq 2$ and $\moy{A_1B_0C_1+A_0B_1C_0+A_0B_0C_1-A_1B_1C_0} \leq 2$, which are facets of the `Class 3' type as defined in~\cite{sliwa}.}. Finally, non-signaling correlations---such that their marginal probability distributions when one discards some parties do not depend on the settings of the discarded parties---also form a polytope, bounded by $\max(|I^{22}|,|J^{22}|) \leq 1$; in particular, a non-signaling correlation such that\footnote{More precisely, $P^{22}(a,b,c|x,y,z) = \frac{1}{8}[1 + (-1)^{a+b+c+xy+yz}]$.} $a+b+c = xy + yz$ (mod 2), which can easily be realized with two Popescu-Rohrlich (PR) boxes~\cite{PRbox} (one shared by Alice and Bob, one shared by Bob and Charlie~\cite{barrett_etal_PRA05}), reaches the values $I^{22}=J^{22}=1$.

\subsubsection{Scenario with one input, four outputs for Bob}
\label{subsubsec_ineq_14}

We now consider the case where Bob has only 1 possible input and 4 possible outputs, while Alice and Charlie still have binary inputs and outputs; this could in practice correspond to the case where Bob performs a complete Bell state measurement, in an entanglement swapping experiment---see section~\ref{subsec_PQ_14} below.

Let us denote Bob's outputs by two bits\footnote{To clarify the notations, note that we use superscripts on Bob's output bits $b^0$ and $b^1$, to distinguish the case where they form one single output (${\boldsymbol b} = b^0 b^1$), from the previous case where $b_0$ and $b_1$ were Bob's outputs for two different inputs ($y=0$ and 1, resp.). Note also that since Bob has only one possible input $y$, we do not need to specify it in $P^{14}(a,b^0 b^1,c|x,z)$.} ${\boldsymbol b} = b^0 b^1 = 00,01,10$ or 11, and by $P^{14}(a,b^0 b^1,c|x,z)$ the correlation shared by Alice, Bob and Charlie. Similarly to the previous case, we now define the tripartite correlation terms
\ba
\moy{A_x B^y C_z}_{P^{14}} &=& \sum_{a,b^0b^1,c} \, (-1)^{a + b^y + c} \, P^{14} ( a, b^0b^1, c | x, z ) \nonumber
\ea
and the linear combinations $I^{14}$ and $J^{14}$ as follows:
\ba
I^{14} &=& \frac{1}{4} \sum_{x,z=0,1} \ \moy{A_x B^0 C_z}_{P^{14}} \, , \label{def_I_14} \\
J^{14} &=& \frac{1}{4} \sum_{x,z=0,1} \ (-1)^{x+z} \moy{A_x B^1 C_z}_{P^{14}} \, . \label{def_J_14}
\ea

As in the previous case, if $P^{14}$ is bilocal, then the following nonlinear inequality necessarily holds:
\begin{equation}
 \sqrt{|I^{14}|} + \sqrt{|J^{14}|} \ \leq \ 1 \,. \label{ineq_14}
\end{equation}
Note that this implies in particular the inequality previously derived in~\cite{biloc1_PRL} (see appendix~\ref{app_previous_ineq}).

\begin{proof}
We show that inequality~(\ref{ineq_14}) can directly be derived from~(\ref{ineq_22}). Indeed, from the correlation $P^{14}$, the three parties can obtain a correlation $P^{22}(a,b,c|x,y,z)$, with now binary inputs and outputs for Bob, if, for a given input $y \in \{0,1\}$, Bob simply outputs the corresponding bit $b^y$. Formally:
\ba
P^{22}(a,b,c|x,y,z) &=& P^{14}(a,b^y=b,c|x,z) \nonumber \\
&=& \sum_{b^0,b^1} \delta_{b,b^y} P^{14}(a,b^0 b^1,c|x,z) \,. \qquad \label{def_P22_from_P14}
\ea
One can easily check that for the correlation $P^{22}$ thus obtained, $\moy{A_x B_y C_z}_{P^{22}} = \moy{A_x B^y C_z}_{P^{14}}$ and the values of $I^{22}$ and $J^{22}$ as defined in~(\ref{def_I_22}--\ref{def_J_22}) coincide with the values of $I^{14}$ and $J^{14}$ obtained from~(\ref{def_I_14}--\ref{def_J_14}).

Suppose now that $P^{14}$ is bilocal. Since the processing from $P^{14}$ to $P^{22}$ is made locally by Bob, then $P^{22}$ is also bilocal, and therefore it satisfies~(\ref{ineq_22}). Since $I^{22} = I^{14}$ and $J^{22} = J^{14}$, then~(\ref{ineq_14}) also holds.
\end{proof}

The projection of the correlation space onto the $(I^{14},J^{14})$ plane can also be seen on Figure~\ref{fig_IJ_plane_22_14}. As before, the bilocal inequality~(\ref{ineq_14}) is tight in this plane. On the other hand, local correlations satisfy the Bell inequalities $|I^{14}|+|J^{14}| \leq 1$. Interestingly, it turns out that non-signaling correlations in this scenario are now also bounded by\footnote{For instance, the non-signaling assumption implies $I^{14}+J^{14} = 1 - \sum P^{14}(a,b^0b^1,c|x,z) \leq 1$, where the sum is over all indices $x,z,a,b^0b^1,c$ such that $b^0 \neq a \oplus c$ and $b^1 \neq a \oplus c \oplus x \oplus z$, and where $\oplus$ denotes the addition modulo 2.} $|I^{14}|+|J^{14}| \leq 1$.

\subsubsection{On the difference between the case with 1 input / 4 outputs \\ and the case with 2 inputs / 2 outputs for Bob}

\label{subsubsec_diff_22_14}

We observe a quite strong similarity between the cases where Bob has binary inputs and outputs (the ``\mbox{22-case}''), and where he has 1 input with 4 possible outputs (the ``\mbox{14-case}''). As we have just seen in the proof of~(\ref{ineq_14}), a scenario of the first kind can easily be obtained from a scenario of the second kind, and constraints on the correlations in the \mbox{22-case} imply constraints in the \mbox{14-case}.
However, this is only a one-way procedure. From a correlation with binary inputs and outputs, one cannot simply obtain a correlation with four outputs: indeed, $b_0$ and $b_1$ in the binary case are incompatible measurement results, and can in general not be outputted simultaneously to define a single 4-valued outcome ${\boldsymbol b} = b_0b_1$.

From a more technical point of view, when looking for explicit (bi-)local decompositions, the definitions of weights $q_{\bar{\alpha}\bar{\beta}\bar{\gamma}}$ or of correlators $e_{\bar{i}\bar{j}\bar{k}}$ for instance will be formally the same, the only difference being in the interpretation: in the \mbox{22-case}, $\beta_0$ and $\beta_1$ are different single-bit outputs, corresponding to different inputs, while in the \mbox{14-case}, $\beta^0$ and $\beta^1$ form a single 2-bit output, for a single input.
Since the outputs $\beta^0$ and $\beta^1$ in the \mbox{14-case} are compatible, the bilocal models are more constrained in that case than in the \mbox{22-case}: correlators of the form $e_{\bar{i},11,\bar{k}}$ (with $\bar{i},\bar{k} \neq 11$) are fixed by correlations $P^{14}$, but not by correlations $P^{22}$.
Although our bilocal inequality does not illustrate this fact, it might be the case that more severe constraints on bilocal correlations can be derived in the \mbox{14-case} than in the \mbox{22-case}.
What can be seen however when looking only at the linear combinations $I^{22/14}$ and $J^{22/14}$ is that, as already mentioned, the Bell inequalities of the form $|I^{22}|+|J^{22}| \leq 1$ for local correlations can be violated by non-signaling correlations; but interestingly, their counterpart $|I^{14}|+|J^{14}| \leq 1$ cannot (see Figure~\ref{fig_IJ_plane_22_14}).

\subsubsection{Scenario with one input, three outputs for Bob}
\label{subsubsec_ineq_13}

Let us finally consider the ``\mbox{13-case}'', where Bob has only 1 input and 3 possible outputs (again, Alice and Charlie still have binary inputs and outputs); in practice, this could correspond to an incomplete Bell state measurement---see section~\ref{subsubsec_PQ_13} below.

To compare with the case with four outputs for Bob, we still use two bits to denote Bob's outputs, ${\boldsymbol b} = b^0b^1 = $ 00, 01 or \mbox{\{10 or 11\}}: here, Bob does not distinguish his outcomes 10 and 11, i.e. $[b^0b^1 = 10] \equiv [b^0b^1 = 11]$. We denote by $P^{13}(a,{\boldsymbol b},c|x,z)$ the correlation shared by Alice, Bob and Charlie.

By analogy with the previous cases, let us now define the following tripartite correlators:
\ba
&& \moy{A_x B^0 C_z}_{P^{\!13}} \nonumber \\
&& \quad \ \ = \sum_{a,c} \ (-1)^{a+c} \ \big[ P^{13}(a,00,c|x,z) + P^{13}(a,01,c|x,z) \nonumber \\[-3mm]
&& \hspace{45mm} - P^{13}(a,\{10 \ {\mathrm{or}} \ 11\},c|x,\!z) \big] , \nonumber \\
&& \moy{A_x B^1 C_z}_{P^{\!13}\!,b^0=0} \nonumber \\
&& \quad \ \ = \sum_{a,c} \ (-1)^{a+c} \ \big[ P^{13}(a,00,c|x,z) - P^{13}(a,01,c|x,z) \big] , \nonumber
\ea
and, in a similar way again as before, the following linear combinations:
\ba
I^{13} &=& \frac{1}{4} \sum_{x,z=0,1} \ \moy{A_x B^0 C_z}_{P^{13}} \, , \label{def_I_13} \\
J^{13} &=& \frac{1}{4} \sum_{x,z=0,1} \ (-1)^{x+z} \moy{A_x B^1 C_z}_{P^{13},b^0=0} \, . \label{def_J_13}
\ea

Once again, all bilocal correlations $P^{13}$ necessarily satisfy
\ba
\sqrt{|I^{13}|} + \sqrt{|J^{13}|} \ \leq \ 1 \, , \label{ineq_13}
\ea
and this inequality is tight in the $(I^{13},J^{13})$ plane.

\begin{proof}
We show here that inequality~(\ref{ineq_13}) can directly be derived from~(\ref{ineq_14}). Indeed, from the correlation $P^{13}$, the three parties can obtain a correlation $P^{14}(a,b,c|x,y,z)$, with now four possible outputs for Bob, in the following very simple way: when Bob gets an outcome ${\boldsymbol b} = 0b^1$, he outputs it directly; when he gets the outcome ${\boldsymbol b} =$ \mbox{\{10 or 11\}}, he outputs ${\boldsymbol b} = 10$ or ${\boldsymbol b} = 11$ at random. Formally:
\ba
\left\{
\begin{array}{rcl}
\! P^{14}(a,0b^1,c|x,z) &=& P^{13}(a,0b^1,c|x,z) \\[2mm]
\! P^{14}(a,1b^1,c|x,z) &=& \demi P^{13}(a,\{10 \ {\mathrm{or}} \ 11\},c|x,z) \,.
\end{array}
\right. \qquad
\ea
One can again check that for the correlation $P^{14}$ thus obtained, $\moy{A_x B^0 C_z}_{P^{14}} = \moy{A_x B^0 C_z}_{P^{\!13}}$ and $\moy{A_x B^1 C_z}_{P^{14}} = \moy{A_x B^1 C_z}_{P^{\!13}\!,b^0=0}$; therefore the values of $I^{14}$ and $J^{14}$ as defined in~(\ref{def_I_14}--\ref{def_J_14}) coincide with the values of $I^{13}$ and $J^{13}$ obtained from~(\ref{def_I_13}--\ref{def_J_13}).

Suppose now that $P^{13}$ is bilocal. Then so is $P^{14}$, which therefore satisfies~(\ref{ineq_14}). Since $I^{14} = I^{13}$ and $J^{14} = J^{13}$, then $P^{13}$ satisfies~(\ref{ineq_13}).
\end{proof}

\begin{figure}
\begin{center}
\epsfxsize=7.2cm
\epsfbox{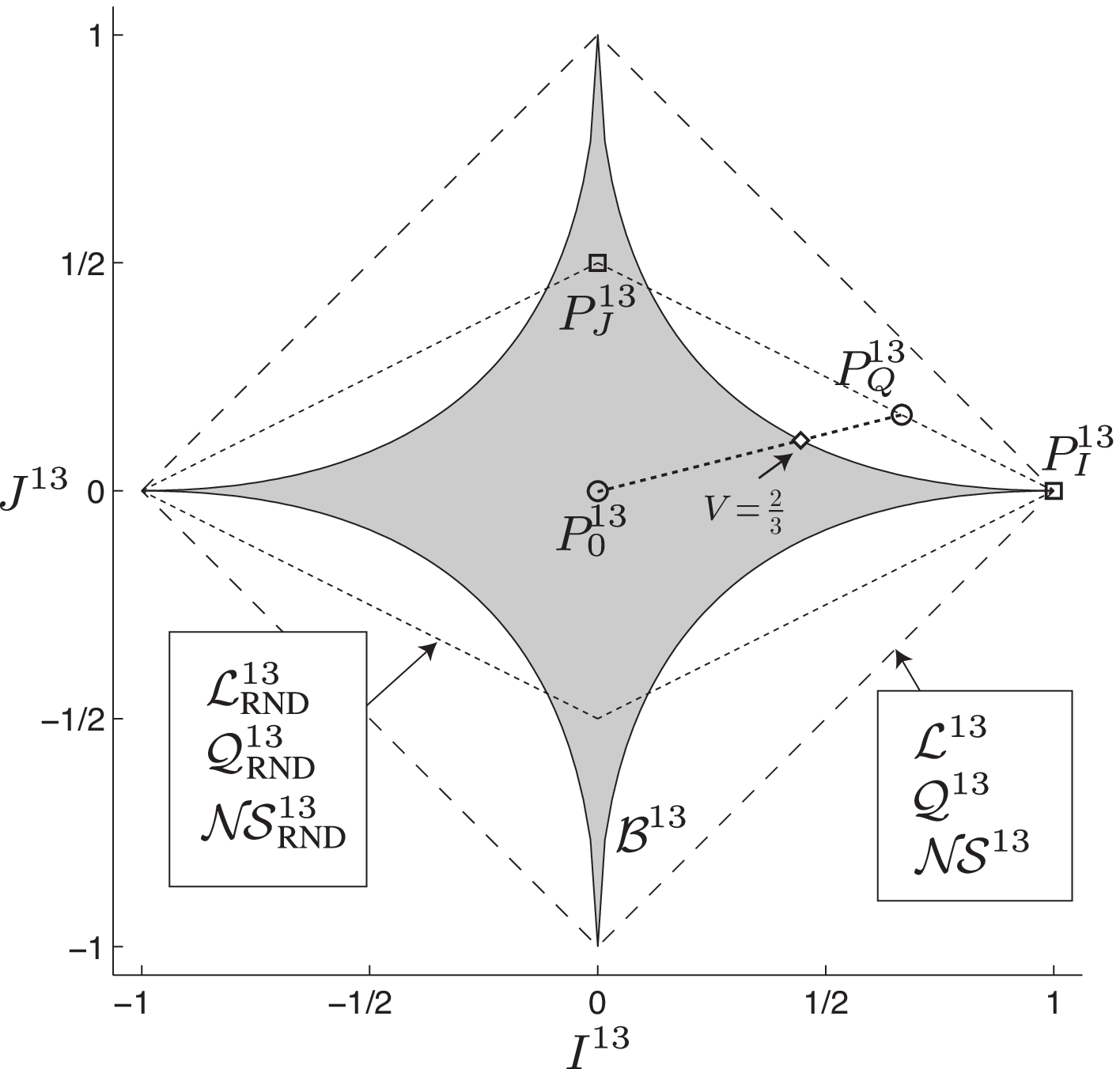}
\caption{Projection of the tripartite correlation space in the $(I^{13},J^{13})$ plane, as defined in (\ref{def_I_13}--\ref{def_J_13}), for the \mbox{``13-case''}. The projections of the bilocal set ${\cal B}^{13}$, of the local polytope ${\cal L}^{13}$, of the quantum set ${\cal Q}^{13}$ and  of the non-signaling polytope ${\cal N\!S}^{13}$ are similar to those of Figure~\ref{fig_IJ_plane_22_14}. When restricting to correlations with a random marginal $\moy{B^0}_{P^{13}}$ for Bob and random bipartite marginals $\moy{A_xC_z}_{P^{13}}$ for Alice-Charlie, the local (${\cal L}^{13}_{\text{RND}}$), quantum (${\cal Q}^{13}_{\text{RND}}$) and non-signaling (${\cal N\!S}^{13}_{\text{RND}}$) sets are instead delimited by $|I^{13}| + 2 |J^{13}| \leq 1$ (dashed diamond).
\\
The restriction of the figure to the dashed diamond can also be understood as a 2-dimensional slice of the correlation space, containing the quantum correlation $P_Q^{13}$~(\ref{eq_PQ_13}), the random correlation $P_0^{13}$, and the bilocal correlations $P_I^{13}$ and $P_J^{13}$, as defined in footnote~\ref{footnote_local_decomp_PQ13}. One can see that the quantum correlation $P_Q^{13} = \frac{2}{3} P_I^{13} + \frac{1}{3} P_J^{13}$ is local, but not bilocal. When adding some noise, it enters the bilocal set for visibilities $V \leq \frac{2}{3}$ (see subsection~\ref{subsubsec_PQ_13}).
}
\label{fig_IJ_plane_13}
\end{center}
\end{figure}

The projection of the correlation space onto the $(I^{13},J^{13})$ plane is shown on Figure~\ref{fig_IJ_plane_13}. As in the \mbox{14-case}, the local and non-signaling sets are delimited in this plane by $|I^{13}|+|J^{13}| \leq 1$, while the bilocal correlations satisfy $\sqrt{|I^{13}|} + \sqrt{|J^{13}|} \leq 1$ (which is tight in this plane: see Table~\ref{table_IJ13} in Appendix~\ref{app_explicit_decomps}).

It is also relevant to restrict ourselves to correlations with a random marginals $\moy{B^0}_{P^{13}}=0$ for Bob (with an obvious notation) and random bipartite marginals $\moy{A_xC_z}_{P^{13}}=0$ for Alice-Charlie, such as the quantum correlation $P_Q^{13}$~(\ref{eq_PQ_13}) studied in subsection~\ref{subsubsec_PQ_13} below. With these additional constraints, the local and non-signaling sets are delimited in the $(I^{13},J^{13})$ plane by\footnote{The non-signaling assumption indeed implies for instance $I^{13}+2J^{13} = 1 + \moy{B^0} - \frac{1}{4} \sum_{x,z} \moy{A_xC_z} - 2 \sum P^{13}(a,0b^1,c|x,z)$, where the last sum is over all indices $x,z,a,b^1,c$ such that $a \neq c$ and $b^1 = x \oplus z$. If Bob's marginal $\moy{B^0}$ and all Alice-Charlie's bipartite marginals $\moy{A_xC_z}$ are zero, it follows that $I^{13}+2J^{13} \leq 1$.} $|I^{13}|+2|J^{13}| \leq 1$; see Figure~\ref{fig_IJ_plane_13}.

\subsubsection{Do our bilocal inequalities fully characterize the bilocal set?}

\label{subsubsec_full_charact}

Note that one can of course also derive many equivalent versions of inequalities~(\ref{ineq_22}),~(\ref{ineq_14}) and~(\ref{ineq_13}), where the inputs and/or outputs are permuted. One may wonder whether these inequalities are enough to delimit the bilocal set, as it is the case for instance with the Clauser-Horne-Shimony-Holt (CHSH) inequality~\cite{chsh} and its equivalent versions for 2 parties with binary inputs and outputs, which---together with trivial inequalities of the form $P \geq 0$---fully characterize the corresponding local set~\cite{fine}. The answer is negative: there exist nonbilocal correlations that satisfy~(\ref{ineq_14}) for instance and all its equivalent versions. Although in most practical cases we study in the next sections, considering inequalities~(\ref{ineq_22}),~(\ref{ineq_14}) or~(\ref{ineq_13}) will be enough to demonstrate non-bilocality, the study of the detection loophole in section~\ref{subsec_det_loop} will provide an example where these are not sufficient; we will then resort to convex relaxation methods.

Fully characterizing the bilocal set, in the scenarios considered here, by a simple list of inequalities remains an open problem.

\section{Quantum violations of bilocality in entanglement swapping experiments}
\label{sec_qviol}

Since John Bell's work in the 1960's~\cite{bell}, it is well understood that his locality assumption can be falsified by quantum correlations. Hence, our bilocality assumption can {\it a fortiori} also be falsified quantum mechanically. Since the latter is a stronger assumption than the former, one may wonder whether it can lead to stronger tests of quantumness; we show now that this is indeed the case.

To justify the use of the bilocality assumption, we consider below scenarios where two independent quantum sources $S_1$ and $S_2$ send particles to Alice and Bob in the state $\varrho_1$, and to Bob and Charlie in the state $\varrho_2$, respectively, so that the overall quantum state is
\ba
\varrho_{ABC} &=& \varrho_1 \otimes \varrho_2 \,. \label{eq_rho_product_state}
\ea
This typically corresponds to entanglement swapping experiments~\cite{zukowski_event-ready-detectors_1993}, as depicted in Figure~\ref{fig_scenarioq}.

We always assume below that Alice and Charlie have binary inputs and outputs. As for Bob, we consider the case where he can perform a full Bell state measurement (he has only one possible input, with four possible outputs), the case where his measurement results group the Bell states two by two, and he can choose among two possible pairings (he has binary inputs and outputs), and the case where his partial Bell state measurement distinghishes two Bell states, but groups the other two together (he has again only one possible input, with now three possible outputs).

In most cases (with the notable exception of subsection~\ref{subsubsec_detLH_Veta} below), our bilocal inequalities derived in the previous section will be sufficient to demonstrate the non-bilocality of the quantum correlations thus obtained.

\subsection{Entanglement swapping experiment \\ with a complete Bell state measurement}

\label{subsec_PQ_14}

We start by considering a typical standard entanglement swapping experiment, where the sources $S_1$ and $S_2$ each produces a singlet state $\ket{\Psi^-}$, and where Bob performs a complete Bell state measurement on the two particles he receives from the two sources; the four possible outcomes ${\boldsymbol b}=b^0b^1=00,01,10$ or $11$ he can obtain correspond to the four Bell states (with standard notations) $\ket{\Phi^+}, \ket{\Phi^-}, \ket{\Psi^+}$ or $\ket{\Psi^-}$, respectively.
Alice and Charlie can each choose a projective measurement (with binary outcomes) to perform on their qubit, described by the observables $\hat{A}_x$ and $\hat{C}_z$ (corresponding to their inputs $x$ and $z$, resp.).

Suppose that Alice and Charlie can measure either $\hat{A}_0 = \hat{C}_0 = (\hat\sigma_{\textsc{z}}+\hat\sigma_{\textsc{x}})/\sqrt{2}$ (for $x,z = 0$) or $\hat{A}_1 = \hat{C}_1 = (\hat\sigma_{\textsc{z}}-\hat\sigma_{\textsc{x}})/\sqrt{2}$ (for $x,z = 1$) on their particle, where $\hat\sigma_{\textsc{z}}$ and $\hat\sigma_{\textsc{x}}$ are the Pauli matrices\footnote{In order not to confuse the notations, we use the fonts $\textsc{x},\textsc{y},\textsc{z}$ for the Pauli matrices $\hat\sigma_{\textsc{x}}, \hat\sigma_{\textsc{y}}$ and $\hat\sigma_{\textsc{z}}$ or for directions on the Bloch sphere, and the fonts $x,y,z$ for Alice, Bob and Charlie's inputs.}. Quantum Mechanics predicts that Alice, Bob and Charlie will observe the following correlation:
\ba
&& \hspace{-.5cm} P_Q^{14}(a,b^0b^1,c|x,z) \nonumber \\[-2mm]
&& \quad = \frac{1}{16} \Big[ 1+(-1)^{a+c} \frac{(-1)^{b^0}+(-1)^{x+z+b^1}}{2} \Big] \,. \label{eq_PQ_14}
\ea

From the definitions~(\ref{def_I_14}--\ref{def_J_14}), one easily obtains
\ba
I^{14}(P_Q^{14}) = J^{14}(P_Q^{14}) \ = \ \frac{1}{2} \,,
\ea
which violates~(\ref{ineq_14}). One can thus conclude that the quantum mechanical correlation $P_Q^{14}$ (\ref{eq_PQ_14}) is non-bilocal.

It turns out however that $P_Q^{14}$ is local\footnote{$P_Q^{14}$ can indeed be decomposed as $P_Q^{14} = \demi P_I^{14} + \demi P_J^{14}$ (see Figure~\ref{fig_IJ_plane_22_14}), with $P_I^{14}(a,b^0b^1,c|x,z) = \frac{1}{16}[1+(-1)^{a+c+b^0}]$ and $P_J^{14}(a,b^0b^1,c|x,z) = \frac{1}{16}[1+(-1)^{x+z+a+c+b^1}]$. $P_I^{14}$ and $P_J^{14}$ are bilocal (and hence, local): they can be obtained from the explicit decompositions of Table~\ref{table_IJ14} in Appendix~\ref{app_explicit_decomps}, for $(I^{14}{=}1,J^{14}{=}0)$ and $(I^{14}{=}0,J^{14}{=}1)$, respectively.
\\
Note that $P_I^{14}$ and $P_J^{14}$, as well as the fully random correlation $P_0^{14}$, are all invariant with respect to the symmetries $(a,b^0,b^1) \leftrightarrow (a{\oplus}1,b^0{\oplus}1,b^1{\oplus}1)$, $(b^0,b^1,c) \leftrightarrow (b^0{\oplus}1,b^1{\oplus}1,c{\oplus}1)$, $(x,b^1) \leftrightarrow (x{\oplus}1,b^1{\oplus}1)$ and $(z,b^1) \leftrightarrow (z{\oplus}1,b^1{\oplus}1)$, which can all be applied bilocally (i.e., independently between $A$--$B$ and $B$--$C$). When applying each of these symmetries with probability $\demi$, any correlation $P^{14}$ (giving values $I^{14},J^{14}$) is projected onto a correlation $P_{\perp}^{14} = I^{14} P_I^{14} + J^{14} P_J^{14} + (1{-}I^{14}{-}J^{14}) P_0^{14}$ on the 2-dimensional slice represented on Figure~\ref{fig_IJ_plane_22_14}. This ``depolarisation'' is similar to that introduced in Ref.~\cite{biloc1_PRL}; similar depolarisation processes can be defined in the other (\mbox{22-} and \mbox{13-}) cases. \label{footnote_local_decomp_PQ14}}.
Like local and non-signaling correlations, quantum correlations are bound to satisfy\footnote{Note that all values of $I^{14},J^{14}$ such that $|I^{14}| + |J^{14}| \leq 1$ can be obtained quantum mechanically; for instance, for Alice and Charlie's measurement settings of the form $\hat{A}_{0/1} = \hat{C}_{0/1} = \cos\theta \hat\sigma_{\textsc{z}} \pm \sin\theta \hat\sigma_{\textsc{x}}$, we get $I^{14}=\cos^2\theta, J^{14}=\sin^2\theta$, and the full line segment $I^{14}+J^{14}=1$ (with $0 \leq I^{14},J^{14} \leq 1$) is recovered. \label{footnote_Q_line_segment}} $|I^{14}| + |J^{14}| \leq 1$ (see Figure~\ref{fig_IJ_plane_22_14}); note that this holds even if the state $\varrho_{ABC}$ does not have the product form~(\ref{eq_rho_product_state}).

\subsubsection{Resistance to noise}

\label{subsubsec_PQV_14}

An interesting figure of merit to quantify the \mbox{non-(bi-)locality} of a correlation is its resistance to noise. One way to model noise\footnote{Note that one could also consider some additional white noise due to the measurement apparatuses, by introducing a visibility $v_P$ for each party $P=A,B$ or $C$. In the case we consider here, the resulting quantum correlation would then depend on the product $V=v_1v_2v_Av_Bv_C$ of the visibilities of each source and measurement apparatus, and our argument would remain unchanged.} is to suppose that each source $S_i$ introduces white noise with probability $1-v_i$, i.e. corresponding to a visibility $v_i$: instead of sending a pure (say, 2-qubit) state $\ket{\psi_i}$, the state it actually sends is
\ba
\varrho_i(v_i) = v_i \, \ket{\psi_i}\bra{\psi_i} + (1-v_i) \, \mathbbm{1}/4 \,. \label{eq_rho_vi}
\ea

In the case we consider here, with maximally entangled states and random marginal probability distributions, the resulting quantum correlation will only depend on the product $V=v_1v_2$ of the visibilities of each source, and is simply given by
\ba
P_Q^{14}(V) \ = \ V P_Q^{14} + (1-V) P_0^{14}, \label{eq_PQV}
\ea
where $P_0^{14}$ is the fully random probability distribution (i.e., $P_0^{14}(a,b^0b^1,c|x,z) = 1/16$ for all $a,b^0b^1,c,x,z$).
The largest visibility for which the correlation $P_Q^{14}(V)$ admits a bilocal decomposition defines the {\it bilocal visibility threshold} $V_{biloc}$, and can be used to quantify the non-bilocality of $P_Q^{14}$: the smallest $V_{biloc}$ is, the more resistant to noise, and hence the more bilocal $P_Q^{14}$ is.

Noting that $I^{14}[P_Q^{14}(V)] = J^{14}[P_Q^{14}(V)] = \frac{1}{2} V$, we conclude from our bilocal inequality~(\ref{ineq_14}) that $P_Q^{14}(V)$ is non-bilocal for all visibilities $V > 50\%$ (see Figure~\ref{fig_IJ_plane_22_14}). On the other hand, for visibilities $V \leq 50\%$, one can find an explicit decomposition that proves that $P_Q^{14}(V)$ is bilocal (see Table~\ref{table_IJ14} in Appendix~\ref{app_explicit_decomps}); our inequality thus detects optimally the resistance to noise of the correlation $P_Q^{14}$, for which $V_{biloc} = 50 \%$.

\medskip

The visibility $V$ can be understood as the visibility of the maximally entangled state that results from the entanglement swapping process. In order to check the nonlocality of that state in the standard locality scenario, one could test the CHSH inequality~\cite{chsh}: this would require the use of different measurement settings for Alice or Charlie, and would require a visibility $V > \frac{1}{\sqrt{2}} \simeq 70.7 \%$ for the CHSH inequality to be violated. Actually, no Bell inequality can be violated (using Von Neumann measurements) for visibilities smaller than $V \simeq 66 \%$~\cite{groth}. Our assumption allows one however to exhibit non-bilocal correlations for visibilities as low as $50 \%$. This illustrates the advantage of the bilocality assumption, which simplifies the requirements for the demonstration of quantumness in entanglement swapping experiments~\cite{matthaeus,entswap_zeilinger}.

\subsubsection{On our choice of measurement settings}

The measurement settings we chose for Alice and Charlie above (and which were already introduced before in~\cite{biloc1_PRL}) are the ones giving the bilocal quantum correlation $P_Q^{14}$~(\ref{eq_PQ_14}) that is the most resistant to noise we could find, i.e. with the smallest bilocal visibility threshold $V_{biloc}$.

These measurement settings were first obtained numerically, following the approach introduced in subsection~\ref{subsec_explicit_models}. Our extensive numerical tests convince us that we have found the optimal settings for the case where the sources send two singlet states, where Alice and Charlie can choose among two projective measurements, and where Bob performs a complete Bell state measurement.
The symmetries of the quantum correlation $P_Q^{14}$~(\ref{eq_PQ_14}) then actually inspired our definitions~(\ref{def_I_14}--\ref{def_J_14}) for $I^{14}$ and $J^{14}$, and the whole analysis of subsection~\ref{subsec_nonlinear_ineq}; interestingly, our bilocal inequality~(\ref{ineq_14}) is sufficient to demonstrate the non-bilocality of $P_Q^{14}(V)$ down to $V=V_{biloc}$.

\subsection{Partial Bell state measurements}

\label{subsec_partialBSM}

An ideal entanglement swapping experiment requires Bob to perform a complete Bell state measurement. This might not be a trivial thing to do; actually, it is known to be impossible to perform this ideal joint measurement with linear quantum optics~\cite{calsamiglia}.

From an experimental perspective, it is therefore interesting to study the consequences of the bilocality assumption in scenarios where Bob does not perform a complete Bell state measurement, but only a partial one. We thus consider below cases where Bob's measurement does not allow him to discriminate the four Bell states, but only subsets of the Bell states.

For instance, he may perform a measurement that allows him to discriminate one Bell states vs the other three. In that case however, we found no advantage with the bilocality assumption, over a test of standard locality (for instance, a test of CHSH between Alice and Charlie, conditioned on Bob having observed the Bell state he can distinguish), even if Bob may have different possible inputs that allow him to choose which of the four Bell states he wants to discriminate: we always found $V_{loc} = V_{biloc} = 1/\sqrt{2}$. Another possibility would be for Bob to distinguish pairs of Bell states. If he can only distinguish two states vs the other two, the correlation shared by the three parties will be bilocal\footnote{This is due to the fact that for a pairwise grouping of the Bell states, Bob's measurement is separable, of the form $\hat\sigma_{\textsc{U}} \otimes \hat\sigma_{\textsc{U}}$, with $\textsc{U} = \textsc{X}, \textsc{Y},$ or $\textsc{Z}$ depending on the grouping (see~\ref{subsubsec_PQ_22}).}; if he can choose among two different pairwise groupings of the same four Bell states (when he has two possible inputs), they will obtain non-bilocal correlations, with a visibility threshold $V_{biloc} = 50\%$ (see subsection~\ref{subsubsec_PQ_22} below); the case where he can choose among the 3 possible pairwise groupings does not provide any advantage over the case with two inputs. Finally, Bob may be able to discriminate two Bell states perfectly, but not to distinguish the other two. In that case, the correlation will again be non-bilocal, with now a visibility threshold $V_{biloc} = 2/3$ (subsection~\ref{subsubsec_PQ_13} below); when Bob has more than one input that allow him to choose which two Bell states he wants to discriminate perfectly, the situation is the same as the previous one, with pairwise groupings.

\subsubsection{Binary inputs and outputs for Bob}

\label{subsubsec_PQ_22}

Let us thus start by considering the case where Bob has two possible inputs, and he wants to distinguish either the $\ket{\Phi^{\pm}}$ vs the $\ket{\Psi^{\pm}}$ Bell states, or the $\ket{\Phi^+/\Psi^+}$ vs the $\ket{\Phi^-/\Psi^-}$ Bell states. That is, he measures either $\hat{B}_0 = \ket{\Phi^+}\bra{\Phi^+} + \ket{\Phi^-}\bra{\Phi^-} - \ket{\Psi^+}\bra{\Psi^+} - \ket{\Psi^-}\bra{\Psi^-} = \hat\sigma_{\textsc{z}} \otimes \hat\sigma_{\textsc{z}}$ or $\hat{B}_1 = \ket{\Phi^+}\bra{\Phi^+} - \ket{\Phi^-}\bra{\Phi^-} + \ket{\Psi^+}\bra{\Psi^+} - \ket{\Psi^-}\bra{\Psi^-} = \hat\sigma_{\textsc{x}} \otimes \hat\sigma_{\textsc{x}}$; note that his measurement is actually a separable measurement.

We still assume that the two sources produce singlet states, and that Alice and Charlie perform the same measurements as before, in subsection~\ref{subsec_PQ_14}: $\hat{A}_{0/1} = \hat{C}_{0/1} = (\hat\sigma_{\textsc{z}} \pm \hat\sigma_{\textsc{x}})/\sqrt{2}$.
In this scenario where all parties have binary inputs and outputs, the quantum correlation shared by the three parties is then
\ba
P_Q^{22}(a,b,c|x,y,z) = \frac{1}{8} \big[ 1 + \demi (-1)^{a+b+c+xy+yz} \big], \quad \label{eq_PQ_22}
\ea
for which, from the definitions~(\ref{def_I_22}--\ref{def_J_22}), one gets
\ba
I^{22}(P_Q^{22}) = J^{22}(P_Q^{22}) \ = \ \frac{1}{2} \,.
\ea
This violates~(\ref{ineq_22}), which proves that $P_Q^{22}$ is not bilocal.

The bilocal visibility threshold can be defined as in the previous section, by considering correlations of the same form as~(\ref{eq_PQV}) (with now $P_0^{22}(a,b,c|x,y,z) = 1/8$ for all $a,b,c,x,y,z$). We find that $P_Q^{22}(V)$ violates inequality~(\ref{ineq_22}) for visibilities $V > 50 \%$, and that it admits a bilocal model for visibilities $V$ (see Table~\ref{table_IJ22} in Appendix~\ref{app_explicit_decomps}), so that again, for $P_Q^{22}$, $V_{biloc} = 50 \%$ (see Figure~\ref{fig_IJ_plane_22_14}).

The practical consequences of this result are however not as interesting as in the \mbox{14-case}. Bob's measurement being separable, we are not considering here an entanglement swapping experiment. In fact, the scenario here amounts to performing two tests of the CHSH inequality, between Alice--Bob and Bob--Charlie. The requirement $V > 50 \%$ simply corresponds to the requirement that at least one of the visibilities $v_i$ of the CHSH tests be larger than $1/\sqrt{2}$; however, this is already a sufficient condition to demonstrate simply Bell nonlocality.

\medskip

For completeness, note that as was $P_Q^{14}$ in the \mbox{14-case}, $P_Q^{22}$ is local\footnote{$P_Q^{22}$ can indeed be decomposed as $P_Q^{22} = \demi P_I^{22} + \demi P_J^{22}$ (see Figure~\ref{fig_IJ_plane_22_14}), with $P_I^{22}(a,b,c|x,y,z) = \frac{1}{8}[1+\delta_{y,0}(-1)^{a+b+c}]$ and $P_J^{22}(a,b,c|x,y,z) = \frac{1}{8}[1+\delta_{y,1}(-1)^{x+z+a+b+c}]$. $P_I^{22}$ and $P_J^{22}$ are bilocal: they can be obtained from the explicit decompositions of Table~\ref{table_IJ22} in Appendix~\ref{app_explicit_decomps}, for $(I^{22}{=}1,J^{22}{=}0)$ and $(I^{22}{=}0,J^{22}{=}1)$, respectively.
\label{footnote_local_decomp_PQ22}}. Like local correlations (but unlike non-signaling ones, here), and as in the \mbox{14-case} once again, quantum correlations actually also satisfy $|I^{22}| + |J^{22}| \leq 1$, which is again tight and holds even if the state $\varrho_{ABC}$ does not have the product form~(\ref{eq_rho_product_state}); this can be seen by expanding the factors of the positive operator $\hat{\cal O}=(\hat A_{+} - \hat B_0 \hat C_{+})^2 + (\hat C_{+} - \hat A_{+} \hat B_0)^2 + (\hat A_{-} - \hat B_1 \hat C_{-})^2 + (\hat C_{-} - \hat A_{-} \hat B_1)^2$, with $\hat A_\pm = \hat A_0 \pm \hat A_1$ and $\hat C_\pm = \hat C_0 \pm \hat C_1$ (and where tensor products are implicit, while identity operators are omitted; e.g., $\hat B_0 \hat C_{+}$ actually stands for $\mathbbm{1} \otimes \hat B_0 \otimes \hat C_{+}$).

\subsubsection{One input, three possible outputs for Bob}

\label{subsubsec_PQ_13}

Another interesting case is when Bob performs a single incomplete Bell state measurement, with now three outcomes. We consider here for instance the case where his outcomes ${\boldsymbol b} = b^0b^1 =$ 00, 01, \mbox{\{10 or 11\}} correspond, respectively, to $\ket{\Phi^+}$, $\ket{\Phi^-}$ and $\ket{\Psi^\pm}$ (which Bob cannot discriminate, as they give the same outcome). Such a measurement can be realized with linear quantum optics~\cite{calsamiglia}, hence the practical motivation for studying this particular case. 

We consider now the following measurements for Alice and Charlie: $\hat{A}_0 = \hat{C}_0 = (\sqrt{2} \, \hat\sigma_{\textsc{z}} + \hat\sigma_{\textsc{x}}) / \sqrt{3}$ and $\hat{A}_1 = \hat{C}_1 = (\sqrt{2} \, \hat\sigma_{\textsc{z}} - \hat\sigma_{\textsc{x}}) / \sqrt{3}$.
The quantum correlation $P_Q^{13}(a,{\boldsymbol b},c|x,z)$ shared by Alice, Bob and Charlie is then
\ba
\left\{
\begin{array}{l}
\! P_Q^{13}(a,0b^1,c|x,z) = \frac{1}{16} \big[ 1 + (-1)^{a+c} \frac{2+(-1)^{x+z+b^{\!1}}}{3} \big] \\[2mm]
\! P_Q^{13}(a,\{10 \ {\mathrm{or}} \ 11\},c|x,z) = \frac{1}{8} \big[ 1 - \frac{2}{3} (-1)^{a+c} \big]
\end{array}
\right. \qquad
\label{eq_PQ_13}
\ea

One easily obtains, from the definitions~(\ref{def_I_13}--\ref{def_J_13}),
\ba
I^{13}(P_Q^{13}) = \frac{2}{3} \,, \quad J^{13}(P_Q^{13}) = \frac{1}{6} \,, \label{eq_I13_J13_PQ}
\ea
which violates~(\ref{ineq_13}), thus proving that $P_Q^{13}$ is not bilocal.

For noisy correlations of the same form as~(\ref{eq_PQV}), with now $P_0^{13}(a,0b^1,c|x,z) = \frac{1}{16}$ and $P_0^{13}(a,\{10 \ {\mathrm{or}} \ 11\},c|x,z) = \frac{1}{8}$, the above values of $I^{13}$ and $J^{13}$ are again simply to be multiplied by $V$. They violate inequality~(\ref{ineq_13}) for all $V > \frac{2}{3}$. For $V \leq \frac{2}{3}$ on the other hand, one can find an explicit bilocal decomposition for $P_Q^{13}(V)$, as in Table~\ref{table_IJ13} of Appendix~\ref{app_explicit_decomps}. Hence, the bilocal visibility threshold of $P_Q^{13}$ is $V_{biloc} = \frac{2}{3}$ (see Figure~\ref{fig_IJ_plane_13}).

Even when a complete Bell state measurement is not possible, the bilocality assumption thus provides an advantage---compared to the standard Bell locality assumption---for practical demonstrations of quamtumness in entanglement swapping experiments.

\medskip

Let us finally note that as in the previous cases, $P_Q^{13}$ is local\footnote{$P_Q^{13}$ can indeed be decomposed as $P_Q^{13} = \frac{2}{3} P_I^{13} + \frac{1}{3} P_J^{13}$ (see Figure~\ref{fig_IJ_plane_13}), with $P_I^{13}(a,0b^1\!\!,c|x,z) = \frac{1}{16}[1+(-1)^{a+c}]$, $P_I^{13}(a,\!\{10 \, {\mathrm{or}} \, 11\},c|x,z)$ $ = \frac{1}{8}[1-(-1)^{a+c}]$ and $P_J^{13}(a,0b^1,c|x,z) = \frac{1}{16}[1+(-1)^{x+z+a+c+b^1}]$, $P_J^{13}(a,\{10 \ {\mathrm{or}} \ 11\},c|x,z) = \frac{1}{8}$. $P_I^{13}$ and $P_J^{13}$ are bilocal: they can be obtained from the explicit decompositions of Table~\ref{table_IJ13} in Appendix~\ref{app_explicit_decomps}, for $(I^{13}{=}1,J^{13}{=}0,K^{13}{=}1,L^{13}{=}M^{13}{=}0)$ and $(I^{13}{=}0,J^{13}{=}\frac{1}{2},K^{13}{=-}1,L^{13}{=}M^{13}{=}0)$, respectively. \label{footnote_local_decomp_PQ13}}. Like local and non-signaling correlations, quantum correlations with random single- and bi-partite marginals (as obtained from singlet states and a partial Bell state measurement for Bob) satisfy\footnote{Note again that all values of $I^{13},J^{13}$ such that $|I^{13}| + 2|J^{13}| \leq 1$ can be obtained quantum mechanically; for instance, for the same measurement settings as in footnote~\ref{footnote_Q_line_segment}, the full line segment $I^{13}+2J^{13}=1$ (with $0 \leq I^{13},2J^{13} \leq 1$) is recovered.} $|I^{13}|+2|J^{13}| \leq 1$; see Figure~\ref{fig_IJ_plane_13}.

\subsection{Resistance to detection inefficiencies}

\label{subsec_det_loop}

Another experimental imperfection that is important to take into account is the fact that Alice, Bob and Charlie's detectors might not be 100\% efficient. In a typical demonstration of nonlocality, this may open the well-known detection loophole~\cite{pearle_detLH}, if the parties post-select their correlations on detected events only.

Restricted classes of local models with independent sources were actually considered before in~\cite{gisin_gisin_02,GHZZ}, and were precisely studied in the context of the detection loophole.
Here we initiate the study of the detection loophole with respect to the general assumption of bilocality, by considering the simplest cases, where only one party has an limited detection efficiency $\eta \in [0,1]$ while other parties have perfect detectors, and the case where both Alice and Charlie have the same detection efficiency $\eta$ while Bob has 100\% efficient detectors. More complex cases, for instance where all parties may have non-prefect detectors, are beyond the scope of this paper, and are left for future work~\cite{denis_detLH_in_preparation}.

In the preliminary study below, we consider again binary inputs and outputs for Alice and Charlie, and a complete Bell state measurement for Bob. We assume that when one party fails to get a conclusive result, they still output a result from their standard set of possible outcomes, either at random or according to a specific strategy---note that when the parties output random results, the situation is the same as for white noise in their measurement apparatus, and as that of imperfect visibilities studied before. The case where a no-detection result is explicitly treated as a different outcome is left for future work; it is an open question, whether this may increase here the resistance to detection inefficiencies.

\subsubsection{Only one party has inefficient detectors}

We start with the case where one party (either Alice, Bob or Charlie) has imperfect detectors with efficiency $\eta$, while the other two have perfectly efficient detectors. Note that in the case of Bob's Bell state measurement, if one of his detectors does not click, he might still get some partial information (such as on a subset of possible Bell states); we assume however that he does not make use of that information---which could possibly lead to a better resistance to detection inefficiencies---and we leave this potential improvement as an open research problem.

We found that the best strategy is for Alice and Charlie to use the same measurement settings as in subsection~\ref{subsec_PQ_14}, which give the correlation $P_Q^{14}$~(\ref{eq_PQ_14}), and for the party with inefficient detectors to output a random result in case of a no-detection event; in that case, similar to that studied previously, the resulting correlation is non-bilocal for $\eta > \eta_{biloc} =  50\%$ (where the {\it bilocal detection efficiency threshold} $\eta_{biloc}$ is defined in a similar way as the bilocal visibility threshold $V_{biloc}$ above); our bilocal inequality~(\ref{ineq_14}) detects optimally its non-bilocality.

\subsubsection{Alice and Charlie both have imperfect detectors ($\eta_A=\eta_C=\eta, \eta_B=1$)}

\label{subsubsec_detLH_Veta}

The second case we consider is that where Alice and Charlie both have imperfect detectors, with the same efficiency $\eta$, while Bob has perfect detectors.
Alice and Charlie still perform the same measurements as before. In case of non-detections, the best strategy is for Alice to output her input directly (i.e., $a=x$), and for Charlie to always output $c=0$. The resulting correlation is found to be bilocal for all $\eta \leq 2/3$; an explicit bilocal decomposition is given in Table~\ref{table_detection_efficiency} of Appendix~\ref{app_explicit_decomps}. On the other hand, convex relaxation techniques mentioned in subsection~\ref{subsubsec_numerical_search} allowed us to establish a numerical upper bound of $\eta = 2/3 + \epsilon$, with $\epsilon \approx 10^{-6}$, above which the correlation is non-bilocal. We conclude that in this case, $\eta_{biloc} =  2/3$.

\begin{figure}
\begin{center}
\epsfxsize=7.2cm
\epsfbox{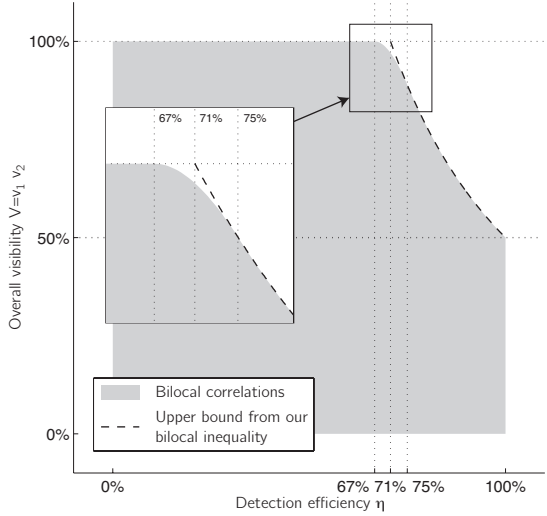}
\caption{(Non-)bilocality of correlations obtained by the strategy described in the main text, in the case where Alice and Charlie's detectors have a detection efficiency $\eta$. The correlations are bilocal for visibilities $V = v_1 v_2 \leq V_{biloc}^{\eta}$, with $V_{biloc}^{\eta}$ depending on $\eta$. As explained in the text, $V_{biloc}^{\eta}$ was estimated by matching lower bounds given by explicit bilocal decompositions (see Table~\ref{table_detection_efficiency} of Appendix~\ref{app_explicit_decomps}) and upper bounds obtained from convex relaxations of the bilocality constraint; our inequality~(\ref{ineq_14}) also gives an upper bound on $V_{biloc}^{\eta}$; this bound was found to be tight only for $\eta \geq \frac{3}{4}$.}
\label{fig_det_vis}
\end{center}
\end{figure}

Taking also into account the noise in the state preparation, as in the previous subsections, the correlation then depends on $\eta$ and on $V = v_1 v_2$. For any fixed value of $\eta$, one can estimate the corresponding bilocal visibility threshold $V_{biloc}^{\eta}$, as shown on Figure~\ref{fig_det_vis}. This was obtained again by comparing a lower bound on $V_{biloc}^{\eta}$ given by the explicit bilocal decomposition of Table~\ref{table_detection_efficiency} in Appendix~\ref{app_explicit_decomps} (to which we refer for analytical expressions for $V_{biloc}^{\eta}$), with a numerical upper bound derived using convex relaxations; the two bounds match again up to $\epsilon \approx 10^{-6}$.

Note that our bilocal inequality~(\ref{ineq_14}) can also be used to obtain an upper bound on $\eta_{biloc}$ and $V_{biloc}^{\eta}$. However, this upper bound on $V_{biloc}^{\eta}$ is found to be tight only for $\eta \geq 3/4$ (see Figure~\ref{fig_det_vis}); and for $V = 1$, inequality~(\ref{ineq_14}) is violated only for $\eta > 1/\sqrt{2}$. This illustrates the fact that our bilocal inequality is not always sufficient to detect the non-bilocality of a correlation (see subsection~\ref{subsubsec_full_charact}).

\subsubsection{Open problems related to the detection loophole}

A more complete study would be necessary to draw any conclusion regarding the advantage of the bilocality assumption compared to the locality assumption, with respect with the detection loophole.

One should in particular consider the case where all three parties have inefficient detectors; note that because their measurements are inherently different, there is no reason to assume that Alice and Charlie should have the same detection efficiencies as Bob; rather, it would be relevant to consider a practical (and incomplete) Bell state measurement and see how Bob's efficiency would compare to that of Alice and Charlie~\cite{denis_detLH_in_preparation}.

Many questions are left open here, including whether using partially entangled or higher-dimensional states may lower the required efficiencies, as it is the case in the standard scenario of Bell nonlocality~\cite{eberhard,detLH_qudits}.

\section{Further issues on quantum non-bilocality}
\label{sec_misc}

We now present some additional results related to the study of quantum non-bilocality, coming back to the case where Bob performs a complete Bell state measurement and where all parties have perfect detectors. We first study the relation between resistance to noise with respect to locality and to bilocality. We then present quantum violations of bilocality using partially entangled qubit states. Finally, we address the question of simulating quantum correlations in a bilocal manner.

\subsection{Trade-off between Bell nonlocality and non-bilocality \\ for quantum correlations}

\label{subsec_pareto}

We have shown that the quantum correlations $P_Q^{14}$~(\ref{eq_PQ_14}), $P_Q^{22}$~(\ref{eq_PQ_22}) and $P_Q^{13}$~(\ref{eq_PQ_13}) we exhibited are not bilocal. They are in fact the most robust to noise we could find in each scenario, and become bilocal for visibilities smaller than $V_{biloc} = 50 \%$ in the first two cases, and $V_{biloc} = 2/3$ in the last case.

However, these correlations were found to be local. In each case, one can also, of course, obtain nonlocal quantum correlations by rotating for instance the measurement settings of Alice and/or Charlie. The nonlocality of such correlations can also be quantified by their resistance to noise: i.e., for a given correlation one can define the {\it local visibility threshold} $V_{loc}$ below which the corresponding noisy correlation of the form~(\ref{eq_PQV}) becomes local.
One may then wonder how the two visibility thresholds, $V_{loc}$ and $V_{biloc}$, behave, one compared to the other, when the measurement settings of Alice and Charlie vary.

\begin{figure}
\begin{center}
\epsfxsize=7.2cm
\epsfbox{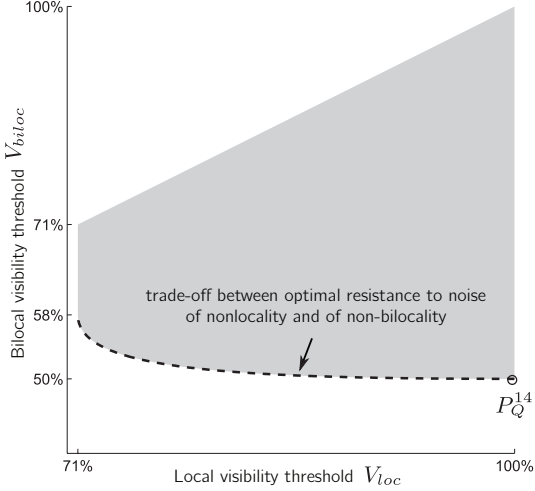}
\caption{Shaded area: local ($V_{loc}$) versus bilocal ($V_{biloc}$) visibility thresholds for quantum correlations obtainable in standard entanglement swapping experiments, with binary inputs and outputs for Alice and Charlie. For the quantum correlation $P_Q^{14}$ of eq.~(\ref{eq_PQ_14}) for instance, on has $(V_{loc}, V_{biloc}) = (1,\demi)$.}
\label{fig_Vloc_Vbiloc_front}
\end{center}
\end{figure}

To illustrate the trade-off between the local and bilocal visibility thresholds, let us consider here the first scenario (the \mbox{14-case}), where Bob performs only one measurement, with four possible outcomes. For the correlation~(\ref{eq_PQ_14}), we had $V_{loc} = 1$ and $V_{biloc} = \demi$. By changing the measurement settings $\hat{A}_0, \hat{A}_1, \hat{C}_0$ and $\hat{C}_1$, while still considering a complete Bell state measurement for Bob, we modify the quantum correlation $P_Q^{14}$ and thus obtain different corresponding pairs of visibility thresholds $(V_{loc}, V_{biloc})$. From a numerical investigation, we obtained the set of pairs $(V_{loc}, V_{biloc})$ represented on Figure~\ref{fig_Vloc_Vbiloc_front}.

The upper boundary illustrates the fact that, obviously, $V_{biloc} \leq V_{loc}$; note that the extreme left point for which $V_{loc} = V_{biloc} = \frac{1}{\sqrt{2}} \simeq 70.7 \%$ can be obtained for instance when one uses the standard settings to test the CHSH inequality~\cite{chsh} between Alice and Charlie (e.g., $\hat\sigma_{\textsc{z}}$ and $\hat\sigma_{\textsc{x}}$ for Alice, $(\hat\sigma_{\textsc{z}} \pm \hat\sigma_{\textsc{x}})/\sqrt{2}$ for Charlie). One can see however that $V_{biloc}$ can be lowered down to $V_{biloc} = 2-\sqrt{2} \simeq 58.6 \%$ while still having $V_{loc} = \frac{1}{\sqrt{2}}$.

The lower boundary is of particular interest, as it expresses the trade-off between optimal resistance to noise of nonlocality and of non-bilocality; it is the result of a multi-objective optimization problem.
This front can for instance be parametrized by considering the following measurement settings:
\ba
& \hat A_0 = \hat C_0 = \cos \theta_0^{\xi} \, \hat\sigma_{\textsc{z}} + \sin \theta_0^{\xi} \, \hat\sigma_{\textsc{x}}\,, \qquad \qquad \qquad \nonumber \\
& \hat A_1 = \hat C_1 = \cos \theta_1^{\xi} \, \hat\sigma_{\textsc{z}} + \sin \theta_1^{\xi} \, \hat\sigma_{\textsc{x}}\,, \qquad \qquad \qquad \nonumber \\
& {\mathrm{with}} \ \theta_i^{\xi} = (-1)^i \frac{\pi}{4} - \xi \frac{\pi}{8}, \ {\mathrm{and}} \ \xi \in [0,1]. \label{eq_settings_pareto}
\ea
A straightforward calculation gives, for the quantum correlation $P_Q^{14}(\xi)$ thus obtained and for the definitions~(\ref{def_I_14}--\ref{def_J_14}), $I^{14} = J^{14} = \frac{1}{4} [1+\cos(\xi \frac{\pi}{4})]$.
For a given visibility $V$, one just has to multiply these values of $I^{14}$ and $J^{14}$ by $V$; the correlation then violates inequality~(\ref{ineq_14}) for all visibilities $V > 1/[1+\cos(\xi \frac{\pi}{4})]$. For visibilities lower than $1/[1+\cos(\xi \frac{\pi}{4})]$ on the other hand, one can find an explicit bilocal decomposition, given in Table~\ref{table_pareto14} of Appendix~\ref{app_explicit_decomps}. Hence, this value is precisely the bilocal visibility threshold $V_{biloc}^{\xi}$ for the quantum correlation obtained with the measurement settings~(\ref{eq_settings_pareto}).

To test for locality, we note that for a tripartite correlation to be local, a necessary condition is that the corresponding bipartite correlation between Alice and Charlie, conditioned on one particular result of Bob, is local~\cite{pironio_lifting}, and therefore (in our case where Alice and Bob have binary inputs and outputs) it must satisfy the CHSH inequality~\cite{chsh}. When conditioned on Bob obtaining $\ket{\Phi^-}$, we find that the value of the CHSH polynomial is $CHSH = 2[\cos(\xi \frac{\pi}{4}) + \sin(\xi \frac{\pi}{4})]$. For a given visibility $V$, the value of $CHSH$ is simply multiplied by $V$, and the CHSH inequality $CHSH \leq 2$ is violated for all $V > 1/[\cos(\xi \frac{\pi}{4}) + \sin(\xi \frac{\pi}{4})]$. On the other hand, one can check that the corresponding correlation is local for visibilities lower than $1/[\cos(\xi \frac{\pi}{4}) + \sin(\xi \frac{\pi}{4})]$; an explicit local decomposition is given in Table~\ref{table_pareto14} of Appendix~\ref{app_explicit_decomps}. Hence, this value is precisely the local visibility threshold $V_{loc}$ for the quantum correlation we consider.

The lower front in Figure~\ref{fig_Vloc_Vbiloc_front} can thus be parametrized as
\ba
(V_{loc}^{\xi}, V_{biloc}^{\xi}) = \Big( \frac{1}{\cos(\xi \frac{\pi}{4}) + \sin(\xi \frac{\pi}{4})}, \frac{1}{1+\cos(\xi \frac{\pi}{4})} \Big) \qquad
\ea
with $\xi \in [0,1]$. For $\xi = 0$ in particular, one gets the most non-bilocal correlations we could find, namely $P_Q^{14}$ as in~(\ref{eq_PQ_14}); for $\xi = 1$, we obtain the point $(V_{loc}^{\xi=1}, V_{biloc}^{\xi=1}) = (\frac{1}{\sqrt{2}}, 2-\sqrt{2})$ that we briefly mentioned before.

We also looked at this relation between the locality and bilocality visibility thresholds in other scenarios and found similar trade-offs, although with different quantitative results. For instance, in the \mbox{22-case} we found numerically that when Alice and Charlie's settings vary, while Bob's measurements are fixed, all correlations for which $V_{loc}=1/\sqrt{2}$ also have $V_{biloc}=1/\sqrt{2}=V_{loc}$; an improvement in $V_{biloc}$ can only be obtained by increasing $V_{loc}$.

\subsection{Non-maximally entangled states}

\label{subsec_nonmaxent}

One might wonder whether our inequality~(\ref{ineq_14}) for bilocality can be violated by non-maximally entangled states, and how non-bilocal the resulting correlations can be.

Let us thus consider the case where the sources $S_1$ and $S_2$ send 2-qubit entangled states of the form
\ba
\ket{\psi_i} = \cos\frac{\theta_i}{2} \ket{01} - \sin\frac{\theta_i}{2} \ket{10} \,,
\ea
with $\theta_i \in [0,\frac{\pi}{2}]$ (and $i=1,2$). We assume that Bob performs a complete Bell state measurement (in the standard Bell basis), and that Alice and Charlie can each choose among two projective measurements to perform on their respective qubits. The tripartite correlators that appear in the definitions~(\ref{def_I_14}--\ref{def_J_14}) of $I^{14}$ and $J^{14}$ are easily found to be
\ba
\moy{A_xB^0C_z}_{P^{14}} &=& \bm{a}_x^{\textsc{z}} \bm{c}_z^{\textsc{z}} \\
\moy{A_xB^1C_z}_{P^{14}} &=& \bm{a}_x^{\textsc{x}} \bm{c}_z^{\textsc{x}} \sin\theta_1 \sin\theta_2
\ea
where $\bm{a}_x^{\textsc{z,x}}$ and $\bm{c}_z^{\textsc{z,x}}$ are the \textsc{z} and \textsc{x} components of the vectors $\vec{\bm{a}}_x$ and $\vec{\bm{c}}_z$ representing Alice and Charlie's measurements in the Bloch sphere, respectively (for inputs $x$ and $z$). We thus obtain
\ba
I^{14} &=\frac{1}{4} & (\bm{a}_0^{\textsc{z}} + \bm{a}_1^{\textsc{z}}) (\bm{c}_0^{\textsc{z}} + \bm{c}_1^{\textsc{z}}) \, , \\
J^{14} &=\frac{1}{4} & (\bm{a}_0^{\textsc{x}} - \bm{a}_1^{\textsc{x}}) (\bm{c}_0^{\textsc{x}} - \bm{c}_1^{\textsc{x}}) \sin\theta_1 \sin\theta_2 \, .
\ea
One can easily see that, in order to maximize $\sqrt{|I|}+\sqrt{|J|}$, the optimal settings of Alice and Charlie should be in the $\textsc{zx}$ plane, symmetric around the $\textsc{z}$ axis. More precisely, we find that the optimal measurements are, for both Alice and Charlie,
\ba
\frac{\hat\sigma_{\textsc{z}}\pm\sqrt{\sin\theta_1 \sin\theta_2}\,\hat\sigma_{\textsc{x}}}{\sqrt{1+\sin\theta_1 \sin\theta_2}} \,,
\ea
leading to
\ba
I^{14} = \frac{1}{1+\sin\theta_1 \sin\theta_2}, \quad J^{14} = \frac{\sin^2\theta_1 \sin^2\theta_2}{1+\sin\theta_1 \sin\theta_2}, \qquad
\ea
and
\ba
\sqrt{|I^{14}|}+\sqrt{|J^{14}|} = \sqrt{1+\sin\theta_1 \sin\theta_2} \,.
\ea
If $\sin\theta_1 \sin\theta_2 > 0$, we have $\sqrt{|I^{14}|}+\sqrt{|J^{14}|} > 1$, which proves that the quantum correlation $P_{\!\theta_{\!1}\!,\theta_{\!2}}^{14}$ thus obtained is non-bilocal.

To study its resistance to noise, one can consider, as before, the case where the source sends noisy states of the form $\rho_i(v_i) = v_i \ket{\psi_i}\bra{\psi_i} + (1-v_i) \mathbbm{1}/4$. Because of the non-random marginals, the noisy correlation does no longer have the simple form of~(\ref{eq_PQV}). However, the values of $I$ and $J$, which only involve tripartite correlation terms, are still simply multiplied by the global visibility $V=v_1v_2$. Inequality~(\ref{ineq_14}) is thus violated for all $V > \frac{1}{1+\sin\theta_1 \sin\theta_2}$.

For all values of $\theta_1,\theta_2, v_1$ and $v_2$, such that $v_1 v_2 \leq \frac{1}{1+\sin\theta_1 \sin\theta_2}$, that we tested, we could find numerically an explicit bilocal decomposition for the correlation $P_{\!\theta_{\!1}\!,\theta_{\!2}}^{14}$. We therefore believe that $\frac{1}{1+\sin\theta_1 \sin\theta_2}$ is precisely its bilocal visibility threshold $V_{biloc}$, and that once again our inequality~(\ref{ineq_14}) detects optimally the non-bilocality of $P_{\!\theta_{\!1}\!,\theta_{\!2}}^{14}$.

Note that as expected, for $\theta_1 = \theta_2 = \frac{\pi}{2}$, we recover the case of maximally entangled states studied in section~\ref{subsec_PQ_14}.

\subsection{Classical simulation of (noisy) entanglement swapping}

\label{subsec_simul}

We have shown in Section~\ref{subsec_PQ_14} that the correlations obtained in an entanglement swapping experiment (with a complete Bell state measurement) can be non-bilocal for visibilities down to $V=v_1 v_2 > 50 \%$. A natural question is whether this visibility threshold $V_{biloc} = 50 \%$ can be lowered, possibly by using more measurement settings on Alice and Charlie's sides.

Studying more complex scenarios, with more settings, rapidly becomes very difficult, because of the nonlinearity and nonconvexity of the bilocality assumption. We could not find so far any scenario where $V_{biloc}$ could be lowered. However, by trying to simulate the noisy entanglement swapping experiment with an explicit bilocal model, one can obtain a lower bound on $V_{biloc}$ for all possible scenarios in which Alice and Charlie perform projective Von Neumann measurements on their qubits, and Bob performs a Bell state measurement.

\subsubsection{A fully bilocal model for a visibility $V = 25\%$}

\label{subsubsec_simul_werner}

We present here a model that reproduces with a visibility $V = 25\%$ all correlations obtained in a standard entanglement swapping experiment, where Alice and Charlie perform Von Neumann measurements on their qubits. The model is inspired by Werner's model~\cite{werner89} which reproduces the noisy singlet state (so-called Werner state) correlations for visibilities $v = 50 \%$: intuitively, one can simulate the two singlet states with visibilities $v_1=v_2= 50 \%$, to obtain an overall visibility $V = v_1 v_2 = 25\%$; the only nontrivial question is how to simulate Bob's Bell state measurement.

In Werner's model, Alice and Bob share a random vector $\vec \lambda$, uniformly distributed on the Bloch sphere ${\cal S}^2$. After reception of a measurement setting $\vec{\bm{a}} \in {\cal S}^2$, Alice outputs $A = -\sign(\vec{\bm{a}}\cdot\vec \lambda)$; after reception of a measurement setting $\vec{\bm{b}} \in {\cal S}^2$, Bob outputs $B = \pm1$ with probability $p(B|\vec{\bm{b}},\vec \lambda) = \frac{1+B \, \vec{\bm{b}} \cdot\vec \lambda}{2}$; note that this corresponds precisely to the quantum prediction for the measurement along the direction $\vec{\bm{b}}$ of a qubit in the pure state $\ket{\vec\lambda} \in {\cal S}^2$.

In a similar spirit, we consider the following bilocal model: Alice and Bob share a random vector $\vec \lambda_1$, Bob and Charlie share a random vector $\vec \lambda_2$, both uniformly distributed on the Bloch sphere ${\cal S}^2$. For measurement settings $\vec{\bm{a}}, \vec{\bm{c}} \in {\cal S}^2$, Alice and Charlie output $A = \sign(\vec{\bm{a}}\cdot\vec \lambda_1)$, $C = \sign(\vec{\bm{c}}\cdot\vec \lambda_2)$. As for Bob, in order to simulate his measurement, he outputs the result ${\bm B}=B^0B^1$ of a Bell state measurement on a 2-qubit pure product state $\ket{\vec\lambda_1}\ket{\vec\lambda_2}$, with the probabilities predicted by Quantum Mechanics. We show in Appendix~\ref{app_simul_werner} that this model indeed reproduces the entanglement swapping correlations, with a visibility $V = 25\%$; of course, one can then also simulate smaller visibilities by introducing some additional noise.

We note that there is a significant gap between the upper bound $V = 50\%$ and the lower bound $V = 25\%$ on the bilocal visibility threshold, for any choice of measurement settings. But the situation is quite similar to the case of locality, where Werner's model reproduces the singlet state correlations for a visibility $v = 50 \%$, while the CHSH inequality~\cite{chsh} allows one to demonstrate nonlocality only for $v > 1/\sqrt{2}$. It is known however that there exists a local model that reproduces the Werner state correlations for a visibility $v \simeq 65.95\%$~\cite{groth}; it would be interesting to see if that model could be adapted to a bilocal model for entanglement swapping correlations. The difficulty is to define an adequate simulation of Bob's Bell state measurement, with local variables that do not have (unlike in Werner's model) a straightforward interpretation as quantum states. On the other hand, some inequalities have been found that demonstrate the nonlocality of Werner states for visibilities $v$ slightly smaller than $1/\sqrt{2}$~\cite{vertesi08}. However, these inequalities involve a very large number of measurement settings. It will certainly be very hard to find better inequalities to decrease the visibility threshold in the case of bilocality.

\subsubsection{Simulation of entanglement swapping with communication?}

\label{subsubsec_simul_toner_bacon}

Toner and Bacon~\cite{toner_bacon}, followed by Degorre {\it et al.}~\cite{degorre05}, have shown that the use of one single bit of communication is enough to classically simulate the quantum correlations obtained from Von Neumann measurements on a singlet state. It is quite natural to wonder whether such a result holds in our case, i.e. whether adding some (limited) classical communication can help to simulate the entanglement swapping experiment in a bilocal manner.

We present in Appendix~\ref{app_simul_toner_bacon} a protocol directly inspired from the communication protocol of~\cite{degorre05}, that uses 2 bits of communication. We find that it allows one to simulate the entanglement swapping correlations with a visibility $V = 4/9 \simeq 44.4\%$; this is indeed better than the visibility of 25\% obtained with the previous bilocal model, but this is still a pretty low visibility, which does not even reach the threshold of $50 \%$.

It might however be possible to increase this visibility---and even obtain a perfect simulation with $V = 1$---with a different protocol, possibly using more (but still finite) communication, in addition to bilocal shared randomness. Recent results in this direction indeed look quite promising~\cite{harry}.

\section{On the assumption of independent sources in standard Bell experiments}

\label{sec_triloc}

We finally come back to the justification of our bilocality assumption. As already emphasized in the introduction, a very similar assumption is actually needed in standard Bell tests, namely that the sources of randomness used to choose the measurement settings are independent from the source emitting the states that are being measured (the ``free choice" or ``measurement independence" assumption~\cite{Bell_free_choice,hall_free_will,barrett_gisin}).

To make this connection more precise, consider a standard bipartite Bell experiment, in which a correlation $P(a,b|x,y)$ is observed. Bell's local causality assumption writes
\ba
P(a,b|x,y) = \! \int \! \mathrm{d} \lambda \, \rho(\lambda) \, P(a|x,\lambda)P(b|y,\lambda) . \quad \label{eq_locality2}
\ea
Assume that the random choice of Alice's setting depends on the hidden state $\lambda_1$ of her random number generator, so that $x$ is chosen with probability $P(x|\lambda_1)$, and $\lambda_1$ follows the distribution $\rho_1(\lambda_1)$; similarly, assume that Bob's setting $y$ is chosen with probability $P(y|\lambda_2)$, where the hidden state $\lambda_2$ of his random number generator follows the distribution $\rho_2(\lambda_2)$. The assumption that the settings can be freely chosen implies that $\lambda_1$, $\lambda_2$ and $\lambda$ must be independent. Together with the local causality assumption, the overall probability distribution $P(a,b,x,y)$ then writes
\ba
P(a,b,x,y) &=& P(a,b|x,y) P(x,y) \nonumber \\
&=& \! \int\!\!\!\!\!\int\!\!\!\!\!\int \! \mathrm{d} \lambda_1 \, \mathrm{d} \lambda \, \mathrm{d} \lambda_2 \, \rho_1(\lambda_1) \, \rho(\lambda) \, \rho_2(\lambda_2) \nonumber \\[-1mm]
&& \hspace{4mm} P(x|\lambda_1) P(a|x,\lambda) P(b|y,\lambda) P(y|\lambda_2) . \qquad \label{eq_loc_abxy}
\ea

\begin{figure*}
\begin{center}
\epsfbox{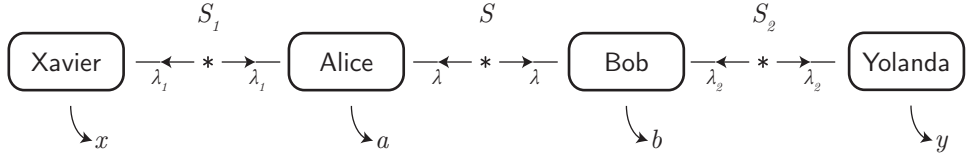}
\caption{A trilocal scenario, with four parties returning outputs $x,a,b,y$ (here they do not receive any input). Under the assumption that $P(x,y)=P(x)P(y)>0$, the 4-partite correlation $P(x,a,b,y)$ is trilocal if and only if the corresponding bipartite correlation $P(a,b|x,y)$ is local.}
\label{fig_scenario_qrng}
\end{center}
\end{figure*}

Let us now compare this situation with the four-partite experiment depicted on Figure~\ref{fig_scenario_qrng}, in which Xavier and Alice receive particles from a source $S_1$, Alice and Bob receive particles from a source $S$, while Bob and Yolanda receive particles from a source $S_2$. The four parties perform some fixed (possibly joint) measurements on their particles, and obtain outputs $x,a,b$ and $y$ respectively. In the spirit of our bilocality assumption, we call {\it trilocal} correlations that can be written in the form
\ba
P(x,a,b,y) &=& \! \int\!\!\!\!\!\int\!\!\!\!\!\int \! \mathrm{d} \lambda_1 \, \mathrm{d} \lambda \, \mathrm{d} \lambda_2 \, \rho_1(\lambda_1) \, \rho(\lambda) \, \rho_2(\lambda_2) \nonumber \\[-1mm]
&& \hspace{3mm} P(x|\lambda_1) P(a|\lambda_1,\lambda) P(b|\lambda,\lambda_2) P(y|\lambda_2) . \qquad \label{eq_trilocality}
\ea
Note the similarities with eq.~(\ref{eq_loc_abxy}).

Assume now that for all $x,y$, the marginal probabilities $P(x,y)$ are $P(x,y) = P(x)P(y) \neq 0$. Under that condition, $P(x,a,b,y)$ is trilocal if and only if the conditional probability distribution $P(a,b|x,y)$, where $x,y$ are interpreted as the inputs of a bipartite scenario, is local.

\begin{proof}

Suppose that $P(x,\!a,\!b,\!y)$ is trilocal, i.e. that it can be decomposed as in~(\ref{eq_trilocality}), and that $P(x,y) = P(x)P(y) \neq 0$ for all $x,y$.
From Bayes' rule, we can write $\rho_1(\lambda_1) P(x|\lambda_1) = \rho_1(\lambda_1|x) P(x)$ and $\rho_2(\lambda_2) P(y|\lambda_2) = \rho_2(\lambda_2|y) P(y)$. Dividing eq.~(\ref{eq_trilocality}) by $P(x)P(y)$, we find that $P(a,b|x,y) = P(x,a,b,y)/[P(x)P(y)]$ is of the form~(\ref{eq_locality2}), with $P(a|x,\lambda) = \int \! \mathrm{d} \lambda_1 \rho_1(\lambda_1|x) P(a|\lambda_1,\lambda)$ and $P(b|y,\lambda) = \int \! \mathrm{d} \lambda_2 \rho_2(\lambda_2|y) P(b|\lambda,\lambda_2)$ (which constitute properly normalised probability distributions). This shows that $P(a,b|x,y)$ is local.

Conversely, suppose that $P(a,b|x,y)$ is local, with a decomposition of the form~(\ref{eq_locality2}), and that $P(x,y)=P(x)P(y)$. Let then $\lambda_1$ and $\lambda_2$ be copies of the variables $x$ and $y$ respectively, so that $P(x|\lambda_1) = \delta_{x,\lambda_1}$ and $P(y|\lambda_2) = \delta_{y,\lambda_2}$.
By writing $P(x) = \sum_{\lambda_1} P(\lambda_1) P(x|\lambda_1)$ and $P(y) = \sum_{\lambda_2} P(\lambda_2) P(y|\lambda_2)$, and using the fact that $P(x|\lambda_1) = \delta_{x,\lambda_1}$ to replace $P(a|x,\lambda)$ by $P(a|\lambda_1,\lambda)$ (and similarly for $P(b|y,\lambda)$), one gets, from the local distribution of $P(a,b|x,y)$, an expression of the form~(\ref{eq_trilocality}) for $P(x,a,b,y) = P(a,b|x,y)P(x)P(y)$: hence, $P(x,a,b,y)$ is trilocal.
\end{proof}

Hence, when Xavier and Yolanda's measurement boxes in Figure~\ref{fig_scenario_qrng} are interpreted as random number generators, which determine Alice and Bob's inputs $x$ and $y$, the four-partite scenario is tantamount to a standard Bell test; the trilocality assumption---which naturally extends our bilocality assumption---is formally equivalent to the assumption that the measurement settings are chosen independently from the source $S$. It is worth stressing that the trilocality assumption can thus be tested without any choice of inputs: there is no need for any additional ``free choice" or ``measurement independence" assumption, as it is already explicitly taken into account in (\ref{eq_trilocality}).

\medskip

To finish off, let us illustrate our claims with an explicit example.
Consider a four-partite scenario as in Figure~\ref{fig_scenario_qrng}, where the sources $S_1$ and $S_2$ send the (separable) states $\varrho_1 = \varrho_2 = \demi \ket{00}\bra{00} + \demi \ket{11}\bra{11}$ to Xavier-Alice and to Bob-Yolanda respectively, and where the source $S$ sends singlet states to Alice-Bob.

Assume that Xavier and Yolanda both measure $\hat X = \hat Y = \sigma_{\textsc{z}}$, that Alice measures $\hat A = \ket{0}\bra{0} \otimes \hat\sigma_{\textsc{z}} + \ket{1}\bra{1} \otimes \hat\sigma_{\textsc{x}}$ while Bob measures $\hat B = \frac{\hat\sigma_{\textsc{z}} + \hat\sigma_{\textsc{x}}}{\sqrt{2}} \otimes \ket{0}\bra{0} + \frac{\hat\sigma_{\textsc{z}} - \hat\sigma_{\textsc{x}}}{\sqrt{2}} \otimes \ket{1}\bra{1}$. The correlation obtained by the four parties is
\ba
P(x,a,b,y) = \frac{1}{16} [ 1 - \frac{1}{\sqrt{2}} (-1)^{a+b+x y} ] \, ,
\ea
leading to
\ba
P(a,b|x,y) = \frac{1}{4} [ 1 - \frac{1}{\sqrt{2}} (-1)^{a+b+x y} ]
\ea
which is precisely the correlation one would get in a Bell test, where the source sends singlet states, where Alice measures either $\hat\sigma_{\textsc{z}}$ or $\hat\sigma_{\textsc{x}}$, and where Bob measures either $\frac{\hat\sigma_{\textsc{z}} + \hat\sigma_{\textsc{x}}}{\sqrt{2}}$ or $\frac{\hat\sigma_{\textsc{z}} - \hat\sigma_{\textsc{x}}}{\sqrt{2}}$.
This can be understood as follows: the measurement of $\hat X$ and $\hat Y$ reveal the inputs $x$ and $y$ of the Bell test; for Alice, measuring $\hat A = \ket{0}\bra{0} \otimes \hat\sigma_{\textsc{z}} + \ket{1}\bra{1} \otimes \hat\sigma_{\textsc{x}}$ precisely amounts to measuring either $\hat\sigma_{\textsc{z}}$ if $x=0$, or $\hat\sigma_{\textsc{x}}$ if $x=1$; similarly for Bob, measuring $\hat B = \frac{\hat\sigma_{\textsc{z}} + \hat\sigma_{\textsc{x}}}{\sqrt{2}} \otimes \ket{0}\bra{0} + \frac{\hat\sigma_{\textsc{z}} - \hat\sigma_{\textsc{x}}}{\sqrt{2}} \otimes \ket{1}\bra{1}$ amounts to measuring either $\frac{\hat\sigma_{\textsc{z}} + \hat\sigma_{\textsc{x}}}{\sqrt{2}}$ if $y=0$, or $\frac{\hat\sigma_{\textsc{z}} - \hat\sigma_{\textsc{x}}}{\sqrt{2}}$ if $y=1$.

\section{Conclusion}
\label{sec_conclusion}

We have developped in this article the study of the bilocality assumption in the context of entanglement swapping experiments. We derived new constraints on bilocal correlations, in the form of nonlinear Bell-type inequalities, for different scenarios; in particular, for the experimentally relevant scenario where only a partial Bell state measurement can be performed. We found in all cases an advantage of the bilocality assumption compared to Bell's standard local causality assumption, as the former lowers the requirements for demonstrating quantumness in entanglement swapping experiments.

A lot of questions are left open. For instance, one could study scenarios with more inputs and outputs, quantum states of higher dimensions, different kinds of measurements performed by the three parties, etc. It is not easy to develop an intuition about which results are to be expected; these questions should motivate further work on the study of bilocal correlations.

Another natural and very interesting problem is to extend our bilocality assumption to more complex topologies of quantum networks with independent sources. While it is straightforward to formulate the assumption of independent sources for the hidden states $\lambda$ (the {\it $N$-locality} assumption) in a similar form as eq.~(\ref{eq_bilocality}) for instance, translating it into more explicit constraints that can be tested numerically, as we did in subsection~\ref{subsec_explicit_models}, is not trivial; let alone deriving Bell-type inequalities for $N$-local correlations.
For instance, a natural extension to our study would be to consider the case of an $(N{+}1)$-partite linear network with $N$ independent sources generating singlet states, where all $N{-}1$ parties inside the chain would perform some Bell state measurements, and the two parties at both ends of the chain can choose among two possible projective measurements; based on our numerical findings for $N=3$, we conjecture that the correlations thus obtained can be \mbox{non-$N$-local} for overall visibilities larger than $V_{N\text{-}loc} = (1/\sqrt{2})^N$ (and that this would be obtained by alternating the horizontal/vertical and the diagonal bases for the measurements of each party), although proving it remains an open problem.

\begin{figure}[b]
\begin{center}
\epsfxsize=5cm
\epsfbox{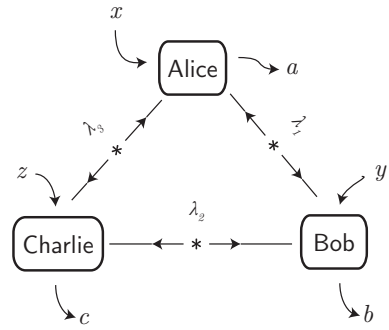}
\caption{A 3-locality scenario, where three parties receive states from 3 sources, forming a closed loop. Characterizing the (non)-3-locality of such a scenario remains an open problem.}
\label{fig_triangle}
\end{center}
\end{figure}

Among the various other network topologies worth investigating in the context of $N$-locality, one is especially intriguing. Consider a simple triangle with Alice, Bob and Charlie at the three vertices (see Figure~\ref{fig_triangle}). In the quantum scenario each of the three edges holds an entangled qubit pair source; the three sources are assumed to be independent, hence the global quantum state is a product of the form $\rho_{AB}\otimes\rho_{BC}\otimes\rho_{AC}$. In the corresponding 3-locality scenario, Alice's output depends on the hidden states ($\lambda_1,\lambda_3$), Bob's on ($\lambda_1,\lambda_2$), Charlie's on ($\lambda_2,\lambda_3$), and the distribution of the three independent states $\lambda_i$ factorizes. One can thus again easily formulate an adequate 3-locality assumption for this scenario in terms of general states $\lambda_i$; however, it is unclear if it is possible to discretize these as we did in subsection~\ref{subsec_explicit_models}, in order to derive more convenient 3-locality constraints. Even in the case without any input, characterizing the (non)-3-locality in this triangle configuration seems challenging.

Studying more deeply the implications of the independent sources assumption will lead to a better understanding of the nonlocality that quantum networks can exhibit, and of how powerful they can be---compared to classical ressources---to perform information processing tasks. We also expect such studies to lead to new applications, fully exploiting the \mbox{non-$N$-locality} of Quantum Mechanics.

\section{Acknowledgments}

We acknowledge Jean-Daniel Bancal, Nicolas Brunner and Yeong-Cherng Liang for fruitful discussions. This work was supported by a University of Queensland Postdoctoral Research Fellowship, by the European ERC-AG QORE, the Swiss NCCR-QSIT, the European EU FP7 QCS project, the CHIST-ERA DIQIP project, the Interuniversity Attraction Poles Photonics@be Programme (Belgian Science Policy), and the Brussels-Capital Region through a BB2B Grant.

\appendix

\section{Topology of the bilocal set}

\label{app_topology}

We prove here that although the bilocal set is non-convex, it keeps some weaker properties of the local set: it is connected, and its restriction to subspaces where the marginal of Alice (or Charlie) is fixed is star-convex (which is not the case however for the whole bilocal set). These properties can for instance be observed on Figures~\ref{fig_IJ_plane_22_14} and~\ref{fig_IJ_plane_13}.

\subsection{Connectedness}

Consider a bilocal correlation $P(a,\!b,\!c|x,\!y,\!z)$, with a bilocal decomposition in terms of local variables $\lambda_1, \lambda_2$, and define the correlation $P_{\xi}$ as follows: each party outputs a result according to the probability distribution $P$ (i.e., according to $P(a|x,\!\lambda_1)$, $P(b|y,\!\lambda_1,\!\lambda_2)$ and $P(c|z,\!\lambda_2)$ resp.) with probability $\xi$, and produces a random output with probability $1-\xi$. The transformation from $P$ to $P_{\xi}$ is made locally (and independently) by each party; therefore, $P_{\xi}$ remains bilocal. For $\xi \in [0,1]$, $P_{\xi}$ follows a continuous path in the bilocal set, from $P=P_{\xi=1}$ to $P_0 = P_{\xi=0}$, where $P_0$ is the fully random probability distribution. All bilocal correlations are thus connected to $P_0$; it follows that the bilocal set is connected.

\subsection{Weak star-convexity (in certain subspaces)}

Let us consider a subspace of $\mathcal{B}_{P(a|x)} \subset \mathcal{B}$ where Alice's marginal probability distribution $P(a|x)$ is fixed. Then $\mathcal{B}_{P(a|x)}$ is star-convex, meaning there exists a point $P_{\!\star} \in \mathcal{B}_{P(a|x)}$ such that the whole line segment between any $P \in \mathcal{B}_{P(a|x)}$ and $P_{\!\star}$ is in $\mathcal{B}_{P(a|x)}$.

\begin{proof}

Let $P_{\!\star}$ be a product correlation of the form $P_{\!\star}(a,\!b,\!c|x,\!y,\!z) = P_{\!\star}(a|x)P_{\!\star}(b|y)P_{\!\star}(c|z)$, with $P_{\!\star}(a|x) = P(a|x)$, while $P_{\!\star}(b|y)$ and $P_{\!\star}(c|z)$ are arbitrary. Clearly, $P_{\!\star} \in \mathcal{B}_{P(a|x)}$.

Consider another correlation $P \in \mathcal{B}_{P(a|x)}$, with a bilocal decomposition in terms of local variables $\lambda_1, \lambda_2$, and let us provide Bob and Charlie with an additional random bit $\ell$, such that $p(\ell=1) = 1-p(\ell=0) = \xi \in [0,1]$. Define now the correlation $P_{\xi}$ as follows: Alice always outputs a result according to $P(a|x,\lambda_1)$; if $\ell=1$, Bob and Charlie output a result according to the probability distributions $P(b|y,\lambda_1,\lambda_2)$ and $P(c|z,\lambda_2)$, resp., while if $\ell=0$ they output a result according to the probability distributions $P_{\!\star}(b|y)$ and $P_{\!\star}(c|z)$.
One can easily check that $P_{\xi} = \xi P + (1-\xi) P_{\!\star}$. Clearly, $P_{\xi}$ is also bilocal, and Alice's marginal satisfies $P_\xi(a|x) = P(a|x)$: hence, $P_\xi \in \mathcal{B}_{P(a|x)}$. For $\xi \in [0,1]$, the full line segment between $P$ and $P_{\!\star}$ in thus in $\mathcal{B}_{P(a|x)}$, which shows that $\mathcal{B}_{P(a|x)}$ is star-convex, for the vantage point $P_{\!\star}$.
\end{proof}

By symmetry, any restriction $\mathcal{B}_{P(c|z)}$ of the bilocal set to a subspace where Charlie's marginal probability distribution is fixed, is also star-convex.

The star-convexity property does not however extend to the full bilocal set. Consider indeed two bilocal correlations $P$ and $P'$, such that both $P(a|x) \neq P'(a|x)$ and $P(c|z) \neq P'(c|z)$, and a mixture $P_\xi = \xi P + (1-\xi) P'$ (with again $\xi \in [0,1]$). Noting that all bilocal correlations necessarily satisfy $P(a,\!c|x,\!z) = P(a|x) P(c|z)$, one gets
\ba
P_\xi(a,\!c|x,\!z) &=& \xi P(a,\!c|x,\!z) + (1-\xi) P'(a,\!c|x,\!z) \nonumber \\
&=& \xi P(a|x) P(c|z) + (1-\xi) P'(a|x) P'(c|z) \nonumber \\
P_\xi(a|x) &=& \xi P(a|x) + (1-\xi) P'(a|x) \nonumber \\
P_\xi(c|z) &=& \xi P(c|z) + (1-\xi) P'(c|z) \, . \nonumber
\ea
and therefore
\ba
&& \hspace{-.3cm} P_\xi(a,\!c|x,\!z) - P_\xi(a|x) P_\xi(c|z) \nonumber \\
&& = \xi (1-\xi) [P(a|x) - P'(a|x)] [P(c|z) - P'(c|z)] \, . \nonumber
\ea
For $0 < \xi < 1$, one thus has $P_\xi(a,\!c|x,\!z) \neq P_\xi(a|x)P_\xi(c|z)$, and therefore $P_\xi$ is not bilocal.
For all bilocal correlation $P$, there thus exists a bilocal correlation to which it is not connected by a line segment of bilocal correlations: this proves that the whole bilocal set is not star-convex (and therefore, in particular, that it is non-convex).

\section{Explicit local and bilocal decompositions}

\label{app_explicit_decomps}

We give in Tables~\ref{table_IJ22} to~\ref{table_pareto14} explicit (bi-)local decomposition for various correlations analyzed in the main text. Exhibiting these decompositions allows us to prove, precisely, that the correlation they define is (bi-)local.
These explicit decompositions are given here in the correlators representation, introduced in subsection~\ref{subsection_correlators}.

\medskip

\begin{table*}

\epsfbox{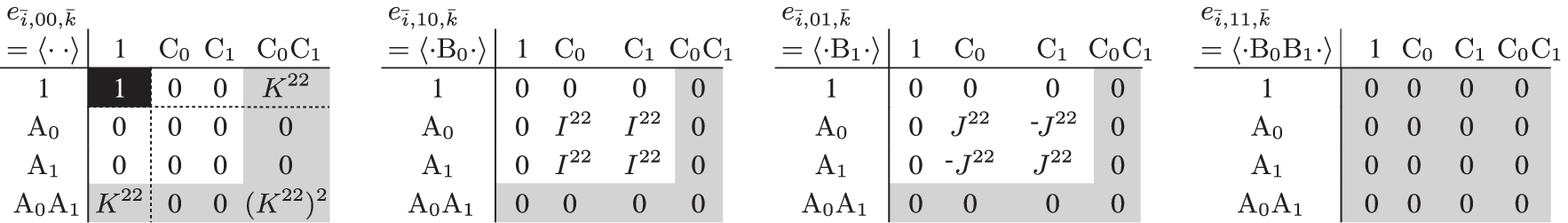}
\caption{Any values of $I^{22}$ and $J^{22}$ such that $\sqrt{|I^{22}|}+\sqrt{|J^{22}|} \leq 1$ can be obtained by a bilocal correlation $P^{22}$, for instance by the one defined by the above explicit decomposition, with $K^{22} = \sqrt{|I^{22}|}-\sqrt{|J^{22}|}$. For this decomposition, the constraint $\sqrt{|I^{22}|}+\sqrt{|J^{22}|} \leq 1$ comes from the non-negativity condition~(\ref{eq_constr_pos_eijk_app}).
\\
For $I^{22} = J^{22} = \frac{1}{2} V$ (and hence $K^{22} = 0$), the table gives a bilocal decomposition for $P_Q^{22}(V)$ (see section~\ref{subsubsec_PQ_22}), valid for $V \in [0,\demi]$.}
\label{table_IJ22}

\vspace{5mm}

\epsfbox{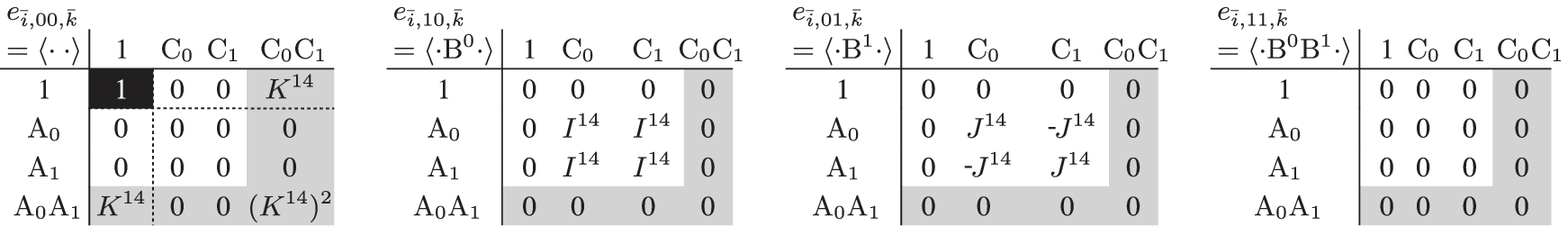}
\caption{Similarly as in the \mbox{22-case} of Table~\ref{table_IJ22}, any values of $I^{14}$ and $J^{14}$ such that $\sqrt{|I^{14}|}+\sqrt{|J^{14}|} \leq 1$ can be obtained by a bilocal correlation $P^{14}$, for instance by the one defined by the above explicit decomposition, with $K^{14} = \sqrt{|I^{14}|}-\sqrt{|J^{14}|}$. Note the strong similarities of this decomposition with that of Table~\ref{table_IJ22}; the main difference being that the correlators $e_{\bar{i},11,\bar{k}}$ with $\bar{i} \neq 11$ and $\bar{k} \neq 11$ (in the fourth sub-table) are no longer internal degrees of freedom of the decomposition; we now display them in non-shaded cells.
\\
For $I^{14} = J^{14} = \frac{1}{2} V$ (and $K^{14} = 0$), the table gives a bilocal decomposition for $P_Q^{14}(V)$ (section~\ref{subsubsec_PQV_14}), valid for $V \in [0,\demi]$.}
\label{table_IJ14}

\vspace{5mm}

\epsfbox{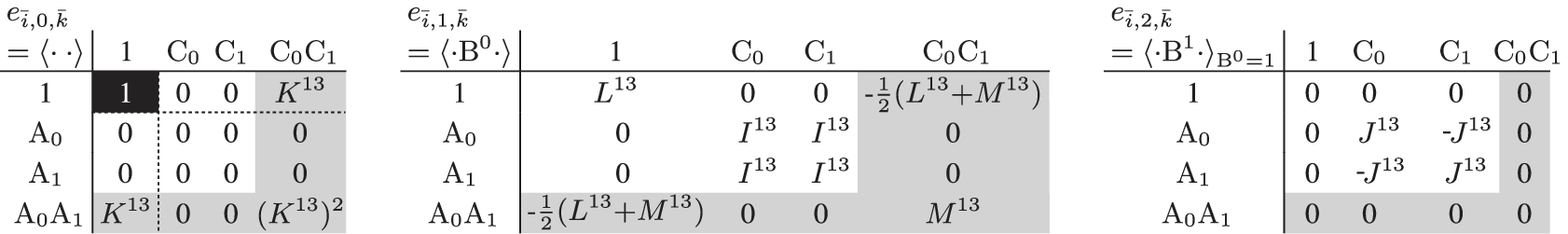}
\caption{Any values of $I^{13}$ and $J^{13}$ such that $\sqrt{|I^{13}|}+\sqrt{|J^{13}|} \leq 1$ can be obtained by a bilocal correlation $P^{13}$, for instance by the one defined by the above explicit decomposition, with $K^{13}, L^{13}$ and $M^{13}$ such that $|L^{13}{+}M^{13}| \leq \demi (1{-}K^{13})^2$, $|L^{13}{-}M^{13}| \leq 1{-}(K^{13})^2$, $4|I^{13}| \leq (1{+}K^{13})^2$ and $4|J^{13}| \leq \demi (1{-}K^{13})^2 + L^{13}{+}M^{13} \leq (1{-}K^{13})^2$.
\\
For $I^{13} = \frac{2}{3}V$, $J^{13} = \frac{1}{6}V$, $K^{13} = \sqrt{|I^{13}|}-\sqrt{|J^{13}|} = \sqrt{\frac{V}{6}}$, $L^{13} = 0$ and $M^{13} = \frac{1}{3}V$, the table gives a bilocal decomposition for $P_Q^{13}(V)$ (section~\ref{subsubsec_PQ_13}), valid for $V \in [0,\frac{2}{3}]$. \\
}
\label{table_IJ13}

\end{table*}

For the case where all parties have binary inputs and outputs (the \mbox{22-case}), we use the definition~(\ref{def_eijk}) for the correlators. A given explicit decomposition will be displayed in the form of four sub-tables (see Table~\ref{table_IJ22}), each of them containing the values of the correlators $e_{\bar{i}\bar{j}\bar{k}}$, for $\bar{j}=00,10,01$ and 11; we refer to eq.~(\ref{eq_eijk_moy}) to clarify the notations on the left column and top row of each sub-table. Let us recall that the correlators $e_{\bar{i}\bar{j}\bar{k}}$ such that $\bar{i} \neq 11, \bar{j} \neq 11$ and $\bar{k} \neq 11$ are fixed by the correlation $P^{22}$ we want to reproduce; we display these in white cells (the number of which is the dimension of the correlation space), except for the constant normalization coefficient $e_{\bar{0}\bar{0}\bar{0}} = 1$, shown in a black cell. On the other hand, the correlators such that $\bar{i} = 11, \bar{j} = 11$ or $\bar{k} = 11$ are internal degrees of freedom of the different possible decompositions of $P^{22}$; these are displayed in shaded cells.

Recall also that a valid decomposition must satisfy the non-negativity constraint~(\ref{eq_constr_pos_eijk}):
\ba
{\mathrm{for \ all}} \ \bar{\alpha},\bar{\beta},\bar{\gamma}, \quad \sum_{\bar{i}\bar{j}\bar{k}} \ (-1)^{\bar{\alpha}\cdot\bar{i}+\bar{\beta}\cdot\bar{j}+\bar{\gamma}\cdot\bar{k}} \ e_{\bar{i}\bar{j}\bar{k}} \geq 0 \, . \label{eq_constr_pos_eijk_app}
\ea
This constraint will delimit the domain of validity of our explicit decompositions.
Finally, the bilocality constraint~(\ref{eq_constr_biloc_eijk}) (i.e. $e_{\bar{i}\bar{0}\bar{k}} = e_{\bar{i}\bar{0}\bar{0}} \, e_{\bar{0}\bar{0}\bar{k}}$) can easily be checked in each case on the first sub-table: the $3 \times 3$ bottom-right sub-table (separated by dashed lines) must be the product of the column on its left with the row above.

\begin{table*}

\setlength{\extrarowheight}{3pt}

\epsfbox{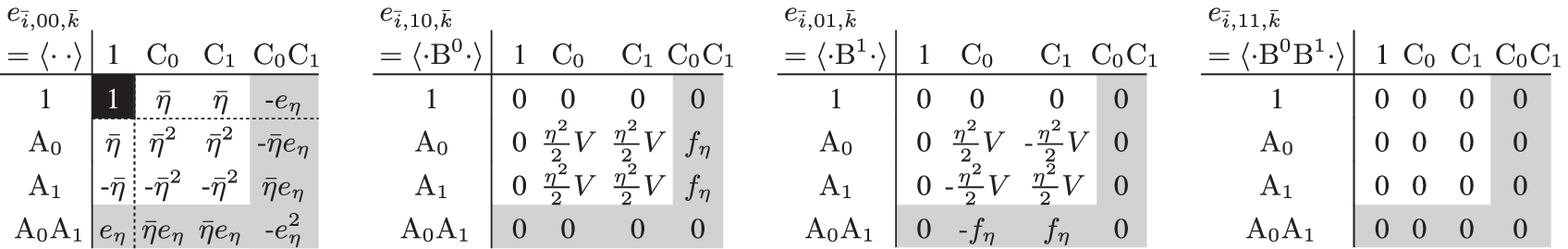}
\caption{Explicit bilocal decomposition for the correlations $P_Q^{14}(V,\eta)$ of section~\ref{subsubsec_detLH_Veta} (shown on Figure~\ref{fig_det_vis}), for imperfect detection efficiencies $\eta$ for Alice and Charlie. We use the notation $\bar\eta=1-\eta$. Three regimes for $\eta$ must be distinguished. For $\eta \geq \frac{3}{4}$, we define $e_{\eta} = 0$, $f_{\eta} = 1{-}\eta$, and the decomposition is valid for all $V \leq V_{biloc}^{\eta} = \frac{1}{2\eta^2}$. For $\frac{2}{3} \leq \eta \leq \frac{3}{4}$, we define $e_{\eta} = 3{-}4\eta$, $f_{\eta} = \min[4(1{-}\eta)^2,4(1{-}\eta)(3\eta{-}2)+V \eta^2]$, and the decomposition is valid for all $V \leq V_{biloc}^{\eta} = \frac{1-e_\eta^2}{2\eta^2} = \frac{4(1{-}\eta)(2\eta{-}1)}{\eta^2}$. Finally, for $\eta \leq \frac{2}{3}$, we define $e_{\eta} = 2\eta{-}1$, $f_{\eta} = V\eta^2$, and the decomposition is valid for all $V \leq V_{biloc}^{\eta} = 1$.}
\label{table_detection_efficiency}

\vspace{5mm}

\epsfbox{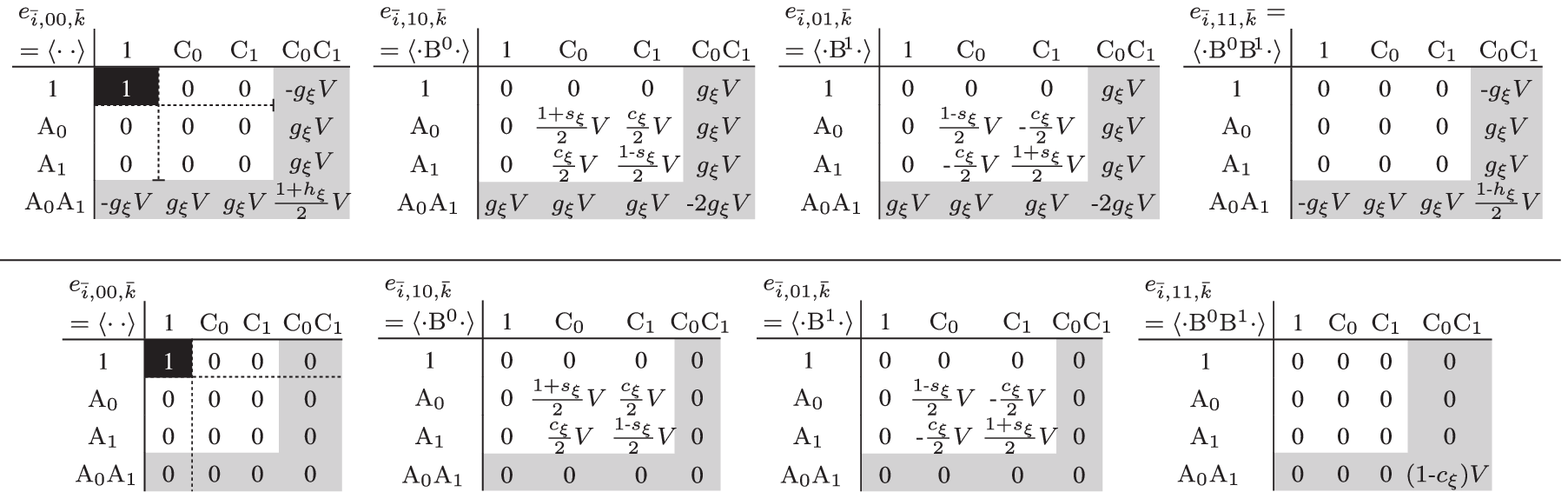}
\caption{Explicit local (top sub-tables) and bilocal (bottom sub-tables) decompositions for the correlation $P_Q^{14}(\xi)$ introduced in subsection~\ref{subsec_pareto} to study the trade-off between resistance to noise of nonlocality and resistance to noise of non-bilocality. We use the notations $c_{\xi} = \cos(\xi \frac{\pi}{4}), s_{\xi} = \sin(\xi \frac{\pi}{4})$, and define $g_{\xi} = \frac{c_{\xi}+s_{\xi}-1}{4}$ and $h_{\xi} = c_{\xi}-s_{\xi}$.
\\
The local decomposition (which is clearly not bilocal: see its first sub-table) is valid, i.e.~(\ref{eq_constr_pos_eijk_app}) is satisfied, for visibilities $V \leq V_{loc} = 1/[c_{\xi}+s_{\xi}]$. The bilocal decomposition is valid for visibilities $V \leq V_{biloc} = 1/[1+c_{\xi}]$. \\
}
\label{table_pareto14}

\end{table*}

As already observed in subsection~\ref{subsubsec_diff_22_14}, the \mbox{14-case}, where Bob has one input and four possible outputs, is quite similar to the \mbox{22-case}. The definitions of the correlators, and the non-negativity and bilocality constraints on these, are formally the same. The only difference is that the correlators $e_{\bar{i},11,\bar{k}}$ with $\bar{i} \neq 11$ and $\bar{k} \neq 11$ are now accessible experimentally, and therefore fixed by the correlation $P^{14}$ we want to reproduce: the explicit decompositions have fewer internal degrees of freedom (see Tables~\ref{table_IJ14} and~\ref{table_detection_efficiency}~--~\ref{table_pareto14}).

\medskip

The \mbox{13-case}, where Bob now has one input and three possible outputs, requires slightly different definitions for the correlators. From a (bi-)local decomposition of a correlation $P^{13}$ in terms of weights $q_{\bar{\alpha}\bar{\beta}\bar{\gamma}}$ (with $\bar{\alpha} = \alpha_0\alpha_1$, $\bar{\gamma} = \gamma_0\gamma_1 =00,01,10$ or 11, and $\bar{\beta} = 00,01$ or \{10 or 11\}), we define in the \mbox{13-case}
\ba
\left\{
\begin{array}{l}
\! e_{\bar{i}0\bar{k}} = \sum_{\bar{\alpha}\bar{\gamma}} (-1)^{\bar{\alpha}\cdot\bar{i}+\bar{\gamma}\cdot\bar{k}} [ q_{\bar{\alpha},00,\bar{\gamma}} + q_{\bar{\alpha},01,\bar{\gamma}} + q_{\bar{\alpha},\{\!10 \mathrm{\, or \, } 1\!1\!\},\bar{\gamma}} ] \, , \\[1mm]
\! e_{\bar{i}1\bar{k}} = \sum_{\bar{\alpha}\bar{\gamma}} (-1)^{\bar{\alpha}\cdot\bar{i}+\bar{\gamma}\cdot\bar{k}} [ q_{\bar{\alpha},00,\bar{\gamma}} + q_{\bar{\alpha},01,\bar{\gamma}} - q_{\bar{\alpha},\{\!10 \mathrm{\, or \, } 1\!1\!\},\bar{\gamma}} ] \, , \\[1mm]
\! e_{\bar{i}2\bar{k}} = \sum_{\bar{\alpha}\bar{\gamma}} (-1)^{\bar{\alpha}\cdot\bar{i}+\bar{\gamma}\cdot\bar{k}} [ q_{\bar{\alpha},00,\bar{\gamma}} - q_{\bar{\alpha},01,\bar{\gamma}} ] \, ,
\end{array}
\right. \nonumber
\ea
which can be inverted into
\ba
\left\{
\begin{array}{l}
\! q_{\bar{\alpha},00,\bar{\gamma}} = 2^{-6} \sum_{\bar{i}\bar{k}} (-1)^{\bar{\alpha}\cdot\bar{i}+\bar{\gamma}\cdot\bar{k}} [ e_{\bar{i}0\bar{k}} + e_{\bar{i}1\bar{k}} + 2e_{\bar{i}2\bar{k}} ] \, , \\[1mm]
\! q_{\bar{\alpha},01,\bar{\gamma}} = 2^{-6} \sum_{\bar{i}\bar{k}} (-1)^{\bar{\alpha}\cdot\bar{i}+\bar{\gamma}\cdot\bar{k}} [ e_{\bar{i}0\bar{k}} + e_{\bar{i}1\bar{k}} - 2e_{\bar{i}2\bar{k}} ] \, , \\[1mm]
\! q_{\bar{\alpha},\{\!10 \mathrm{\, or \, } 1\!1\!\},\bar{\gamma}} = 2^{-6} \sum_{\bar{i}\bar{k}} (-1)^{\bar{\alpha}\cdot\bar{i}+\bar{\gamma}\cdot\bar{k}} [ 2e_{\bar{i}0\bar{k}} - 2e_{\bar{i}1\bar{k}} ] \, .
\end{array}
\right. \nonumber
\ea
With these definitions, one can check that again, all correlators of the form $e_{\bar{i}j\bar{k}}$, with $\bar{i} \neq 11$ and $\bar{k} \neq 11$, are fixed by the correlation $P^{13}$ one wants to reproduce; in particular, one has $e_{00,0,00} = 1$ by normalization, and the correlators $e_{\bar{i}1\bar{k}}$ and $e_{\bar{i}2\bar{k}}$ with $\bar{i},\bar{k} = 10$ or 01 correspond precisely to the tripartite correlation terms $\moy{A_x B^0 C_z}_{P^{\!13}}$ and $\moy{A_x B^1 C_z}_{P^{\!13}\!,b^0=0}$ defined in subsection~\ref{subsubsec_ineq_13}.
Note that for the decomposition to be valid, the weights $q_{\bar{\alpha}\bar{\beta}\bar{\gamma}}$ must be non-negative, which (from the previous equations) imposes some constraints on the correlators $e_{\bar{i}j\bar{k}}$. Finally, the bilocality assumption writes again, in terms of correlators, $e_{\bar{i}0\bar{k}} = e_{\bar{i}0\bar{0}} \, e_{\bar{0}0\bar{k}}$.

In Table~\ref{table_IJ13}, we display the correlators $e_{\bar{i}j\bar{k}}$ in three sub-tables, using similar conventions as before.

\section{Comparison of inequality~(\ref{ineq_14}) \\ with that previously presented in ref.~\cite{biloc1_PRL}}

\label{app_previous_ineq}

Here we show that the bilocal inequality previously derived in~\cite{biloc1_PRL} is actually implied by our inequality~(\ref{ineq_14}), for the case where Bob has one input and 4 possible outputs.

Defining $I_\pm = 2I^{14} \pm 2J^{14}$, inequality~(\ref{ineq_14}) implies
\ba
I_+ \leq 1 + \frac{I_-^2}{4} \, . \label{ineq_IpIm}
\ea

Now, from the definitions~(\ref{def_I_14}--\ref{def_J_14}), we have
\ba
I_+ &=& \sum_{b^0b^1} \ \sum_{x \oplus z = b^0 \oplus b^1} \ P(b^0b^1) \, E_{b^0b^1}(xz) \nonumber \\
I_- &=& \sum_{b^0b^1} \ \sum_{x \oplus z \neq b^0 \oplus b^1} \ P(b^0b^1) \, E_{b^0b^1}(xz) \nonumber
\ea
with
\ba
P(b^0b^1)E_{b^0b^1}(xz) = (-1)^{b^0} \sum_{a,c} (-1)^{a+c} P^{14}(a,b^0 b^1\!,c|x,z) . \nonumber
\ea

One can see that, up to a small change of notations ($b_1 \leftrightarrow b_0 \oplus b_1$), $I_+$ corresponds to $I$, and that $|I_-| \leq 2E$, for $I$ and $E$ as defined in~\cite{biloc1_PRL}. Together with~(\ref{ineq_IpIm}), we thus find that our inequality~(\ref{ineq_14}) implies the one ($I \leq 1+E^2$) previously presented in~\cite{biloc1_PRL}.

It is in fact strictly stronger than the previous inequality: there exist nonbilocal correlations (such as, e.g., the local correlation defined by its weights $q_{00,00,00} = q_{01,01,01} = \demi$, with all its other weights $q_{\bar{\alpha}\bar{\beta}\bar{\gamma}} = 0$) that violate inequality~(\ref{ineq_14}), but which do not violate the inequality of Ref.~\cite{biloc1_PRL}, nor any of its equivalent versions.

\section{Classical simulation \\ of noisy entanglement swapping}

\label{app_simul}

We give in this appendix the details of the simulation of noisy entanglement swapping correlations by bilocal models, as presented in section~\ref{subsec_simul}.

\subsection{Simulation without communication}

\label{app_simul_werner}

Consider the bilocal simulation protocol presented in section~\ref{subsubsec_simul_werner}. To study the correlation $P(A,B^0B^1,C|\vec{\bm{a}}, \vec{\bm{c}})$ it defines, it is sufficient to calculate the average values
\ba
& \hspace{-.3cm} \moy{A}, \moy{C}, \moy{B^0}, \moy{B^1}, \moy{B^0\!B^1}, \nonumber \\
& \hspace{-.3cm} \moy{AC}, \moy{AB^0}, \moy{AB^1}, \moy{AB^0\!B^1}, \moy{B^0C}, \moy{B^1C}, \moy{B^0\!B^1C}, \nonumber \\
& \hspace{-.3cm} \moy{AB^0C}, \moy{AB^1C}, \moy{AB^0B^1C} \,. \label{correls_to_check}
\ea

For ${\bm B} = B^0B^1 \in \{++,+-,-+,--\}$ being the result of a Bell state measurement, corresponding to the outcomes $\ket{\Phi^+}, \ket{\Phi^-}, \ket{\Psi^+}$ and $\ket{\Psi^-}$ respectively, it is convenient to note that the outcome $B^0$ actually corresponds to the measurement of $\hat\sigma_{\textsc{z}} \otimes \hat\sigma_{\textsc{z}}$, the outcome $B^1$ corresponds to the measurement of $\hat\sigma_{\textsc{x}} \otimes \hat\sigma_{\textsc{x}}$, and the product $B^0B^1$ corresponds to the measurement of $-\hat\sigma_{\textsc{y}} \otimes \hat\sigma_{\textsc{y}}$; therefore, the expectation values for a given quantum state $\ket{\vec\lambda_1}\ket{\vec\lambda_2}$ are $\moy{B^0}_{\vec\lambda_1\vec\lambda_2} = \lambda_1^{\textsc{z}}\lambda_2^{\textsc{z}}$, $\moy{B^1}_{\vec\lambda_1\vec\lambda_2} = \lambda_1^{\textsc{x}}\lambda_2^{\textsc{x}}$ and $\moy{B^0B^1}_{\vec\lambda_1\vec\lambda_2} = -\lambda_1^{\textsc{y}}\lambda_2^{\textsc{y}}$ ($\lambda_i^{{\textsc{x}},{\textsc{y}},{\textsc{z}}}$ being the ${\textsc{x}},{\textsc{y}},{\textsc{z}}$ components of $\vec\lambda_i$).

After averaging over $\vec\lambda_1$ and $\vec\lambda_2$, one can easily check that all single- and bi-partite correlators in the first 2 lines of~(\ref{correls_to_check}) vanish, as it is the case for entanglement swapping correlations. To calculate the tripartite correlators, one can show that $\int_{{\cal S}^2} {\mathrm d}\vec\lambda \rho(\vec\lambda) \ \sign(\vec{\bm{a}}\cdot\vec \lambda) \ (\vec \lambda \cdot \vec{\bm{u}}) = \demi \vec{\bm{a}}\cdot\vec{\bm{u}}$ (with $\rho(\vec\lambda) = \frac{1}{4\pi}$ being the uniform distribution of $\vec \lambda$ on the sphere ${\cal S}^2$), so that we get
\ba
& \moy{AB^0C} = \int_{{\cal S}^2} {\mathrm d}\vec\lambda_1 \rho(\vec\lambda_1) \ \sign(\vec{\bm{a}}\cdot\vec \lambda_1) \ \lambda_1^{\textsc{z}} \hspace{2cm} \nonumber \\
& \hspace{2cm} \times \int_{{\cal S}^2} {\mathrm d}\vec\lambda_2 \rho(\vec\lambda_2) \ \sign(\vec{\bm{c}}\cdot\vec \lambda_2) \ \lambda_2^{\textsc{z}} = \frac{1}{4} \, \bm{a}^{\textsc{z}} \, \bm{c}^{\textsc{z}} \, , \nonumber \\
& \moy{AB^1C} = \frac{1}{4} \, \bm{a}^{\textsc{x}} \, \bm{c}^{\textsc{x}} \, , \quad \moy{AB^0\!B^1C} = - \frac{1}{4} \, \bm{a}^{\textsc{y}} \, \bm{c}^{\textsc{y}} \, , \quad \label{eq_app_simul_ABC}
\ea
where $\bm{a}^{\textsc{x,y,z}}$ and $\bm{c}^{\textsc{x,y,z}}$ are the ${\textsc{x}},{\textsc{y}},{\textsc{z}}$ components of the measurement settings $\vec{\bm{a}}$ and $\vec{\bm{c}}$.

Now, for the quantum correlations in an entanglement swapping experiment, one has, precisely,
\ba
& \moy{AB^0C} = \bm{a}^{\textsc{z}} \, \bm{c}^{\textsc{z}} \, , \quad \moy{AB^1C} = \bm{a}^{\textsc{x}} \, \bm{c}^{\textsc{x}} \, , \nonumber \\ & \moy{AB^0\!B^1C} = - \bm{a}^{\textsc{y}} \, \bm{c}^{\textsc{y}} \, . \nonumber
\ea
Hence, our bilocal model reproduces the entanglement swapping correlations with a visibility $V = \frac{1}{4} = 25 \%$.

\subsection{Simulation with 2 bits of communication}

\label{app_simul_toner_bacon}

We have mentioned in Section~\ref{subsubsec_simul_toner_bacon} that one could increase the visibility of the simulation with the help of communication. Inspired from the communication protocol presented in~\cite{degorre05} (Theorem~10) that simulates the singlet state correlations, we slightly modify the previous bilocal model in the following way: instead of starting with $\vec\lambda_1$ and $\vec\lambda_2$ uniformly distributed on the Bloch sphere, we use 1 bit from Alice to Bob, and 1 bit from Charlie to Bob to bias the distributions of $\vec\lambda_1$ and $\vec\lambda_2$ according to (see Theorem~6 of~\cite{degorre05})
\ba
\rho_{\vec{\bm{a}}}(\vec\lambda_1) = \frac{|\vec{\bm{a}} \cdot \vec\lambda_1|}{2\pi}\,, \quad \rho_{\vec{\bm{c}}}(\vec\lambda_2) = \frac{|\vec{\bm{c}} \cdot \vec\lambda_2|}{2\pi} \,. \nonumber
\ea

One can still easily check that the single- and bi-partite correlators in~(\ref{correls_to_check}) vanish, as for the entanglement swapping correlations. The tripartite correlators can be calculated in a similar way as in~(\ref{eq_app_simul_ABC}), with the modified distribution functions $\rho_{\vec{\bm{a}}}(\vec\lambda_1)$ and $\rho_{\vec{\bm{c}}}(\vec\lambda_2)$ above. An easy calculation shows that $\int_{{\cal S}^2} {\mathrm d}\vec\lambda \rho_{\vec{\bm{a}}}(\vec\lambda) \sign(\vec{\bm{a}}\cdot\vec \lambda) (\vec \lambda \cdot \vec{\bm{u}}) = \frac{1}{2\pi} \int_{{\cal S}^2} {\mathrm d}\vec\lambda (\vec{\bm{a}}\cdot\vec \lambda) (\vec \lambda \cdot \vec{\bm{u}}) = \frac{2}{3} \vec{\bm{a}} \cdot \vec{\bm{u}}$, from which we now get
\ba
& \moy{AB^0C} = \frac{4}{9} \, \bm{a}^{\textsc{z}} \, \bm{c}^{\textsc{z}} \, , \ \moy{AB^1C} = \frac{4}{9} \, \bm{a}^{\textsc{x}} \, \bm{c}^{\textsc{x}} \, , \nonumber \\ & \moy{AB^0B^1C} = -\frac{4}{9} \, \bm{a}^{\textsc{y}} \, \bm{c}^{\textsc{y}} \, . \nonumber
\ea
This shows that our bilocal model, augmented by two bits of communication, reproduces the entanglement swapping correlations with a visibility $V = \frac{4}{9} \simeq 44.4 \%$.


\bibliography{bib_bilocality}

%

\end{document}